\documentclass[a4paper,11pt]{article}
\pdfoutput=1 

\usepackage{jcappub} 

\usepackage[T1]{fontenc} 

\usepackage{amsmath}
\usepackage{amsfonts}
\usepackage{amssymb}
\usepackage{graphicx, rotating}
\usepackage{epstopdf}
\usepackage{epsfig}
\usepackage{latexsym}
\usepackage{multirow}
\usepackage{color}
\usepackage[dvipsnames]{xcolor}
\usepackage{slashed}
\usepackage[top=2.8 cm, bottom=2.8 cm, left=3 cm, right=3 cm]{geometry}
\usepackage{amstext} 
\usepackage{array} 

\definecolor{DarkGreen}{rgb}{0.0, 0.2, 0.13}

\pdfoutput=1
\usepackage{listings}
\hypersetup{citecolor={blue}}

\newcolumntype{L}{>{$}l<{$}} 
\newcommand{\ba}{\begin{eqnarray}}
\newcommand{\ea}{\end{eqnarray}}

\newcommand{\be}{\begin{equation}}
\newcommand{\ee}{\end{equation}}

\usepackage{hyperref}
\hypersetup{
colorlinks = true,
linkcolor = black,
citecolor={blue}
}

\title{\boldmath Annual modulations from secular variations: not relaxing DAMA?}

\author[a,b]{Andrea Messina}
\author[a,b]{Marco Nardecchia}
\author[a,b,1]{Stefano Piacentini\note{Corresponding author.}}

\affiliation[a]{Sapienza, Universit\`a di Roma, Italy}
\affiliation[b]{INFN, Sezione di Roma, Italy}

\emailAdd{andrea.messina@uniroma1.it}
\emailAdd{m.nardecchia@uniroma1.it}
\emailAdd{stefano.piacentini@uniroma1.it}

\abstract{In a recent paper~[arXiv:2002.00459], Buttazzo et al. show how the annually modulated rate of the DAMA experiments can be possibly interpreted as an artefact due to the interplay between a time-dependent background and the method to account for it. In this work, we compare this hypothesis against the sinusoidal dark matter signal as proposed by the DAMA collaboration. We produce in a Bayesian approach a quantitative comparison of how much the experimental observations are in support of each hypothesis. 
Our conclusions are that the odds against the hypothesis of a time varying background being responsible for the annual modulation are decreased by a Bayes factor larger than $10^8$ after considering the public available data of the DAMA/NaI and DAMA/LIBRA experiments. 
In this work we also elaborate on general aspects of the analysis procedure.
Indeed, in order to optimise the background subtraction procedure, the DAMA collaboration only considers data-taking cycles with a duration of roughly one year. We argue that any data-taking cycle is informative, and we propose a procedure to include this effect, as well as 
the possibility to include a slowly varying component for the background.}


\arxivnumber{2003.03340}
\begin{document}
\maketitle
\flushbottom

\section{Introduction} 

The results of the DAMA/NaI and DAMA/LIBRA experiments are interpreted by the DAMA collaboration as a strong evidence of the presence of Dark Matter (DM) particles in the galactic halo~\cite{Bernabei:2000qi, DAMA, 0501412, 0804.2741,1308.5109, 1805.10486}. This conclusions derive from the compatibility of the annually modulated rate with a sinusoidal signal, characterized by a phase and a period in agreement with those expected from the interaction of DM particles.

In a recent paper~\cite{Buttazzo:2020bto} it has been discussed the possibility that, due to the analysis procedure followed by the DAMA collaboration, the observed annual modulation could be reproduced by a slowly varying time-dependent background. 
This possibility has been used to interpret the modulation of residuals of the single-hit scintillation rate as a function of time  published by the DAMA collaboration~\cite{Bernabei:2000qi, 0804.2741, 1805.10486}. 

The argument proposed in ref.~\cite{Buttazzo:2020bto} goes as follows: since the residual rate is computed by subtracting the average total rate in every cycle of data-taking, if the background rate is not constant over time this procedure can generate a time modulated rate.
More precisely, in the hypothesis of the actual presence of a DM signal, the total rate $r(t)$ is given by:
\begin{equation}
r(t) = r_0 (t) + A \cos{\left(\frac{2 \pi t}{T} - \phi\right)},
\end{equation}
where $r_0(t)$ is the time-dependent background, $A$ is the amplitude of the oscillating DM signal, $T$ is the period of the oscillation, which is equal to 1 year, and $\phi$ is a phase such that the peak of the oscillation is around the $2^{\text{nd}}$ June, as expected in DM galactic halo models. If $r_0$ is constant over time, then one can choose a time window of width $\Delta$ equal to any multiple of $T$ and average the rate in that window to isolate the background contribution.
Under such an hypothesis, the residuals obtained by subtracting window by window this average are:
\begin{equation}
S(t) \equiv r(t) - \langle r(t)\rangle_{\Delta} = A  \cos{\left(\frac{2 \pi t}{T} - \phi\right)}
\label{eq:St}
\end{equation}
and then the signal is isolated from the background. This is what has been done by the DAMA collaboration, with the time windows $\Delta$ chosen to be of the order of  1 year.

On the other hand, as it has been pointed out,
if the background is slowly varying over time, for example linearly,
the choice of using time windows of the order of the period of the wanted signal can produce residuals with a modulation of the same period.
The possibility of a time-varying background is supported by the fact that in underground detectors, especially in the keV energy range, the features of the background are not completely modeled. In particular in ref.~\cite{Buttazzo:2020bto} explanations due to out-of-equilibrium physical or instrumental effects are considered.
The authors of ref.~\cite{Buttazzo:2020bto} conclude that the available data could not safely exclude the extreme possibility that the modulation in the residual rate is produced by a slowly varying background only. 
Such a conclusion is based on the debatable argument that the DAMA residuals could be fitted by an linearly growing background with a $\chi^2/$~d.o.f~${\simeq1}$.

In ref.~\cite{Krishak:2019hlo} a Bayesian comparison between the cosine and the null hypothesis on the DAMA (2-6)~keVee energy window dataset has been performed, leading to a decisive evidence against the null hypothesis.
The aim of this paper is to perform a thorough comparison of the two hypotheses of a cosine modulation and a slowly varying background, and give a quantitative conclusion on their performance by studying the likelihood ratio, the model complexity, and the posterior probabilities. We highlight that the odds in favor of the cosine modulation hypothesis are decreased by a Bayes factor of the order of $10^8$ with respect to pure linearly varying background hypothesis, when considering the (2-6)~keVee energy window of all the three experimental phases of DAMA.
For the sake of completeness, this conclusion stands out also considering only the likelihood ratio of the hypotheses.  
However, we second the proposal of  ref.~\cite{Buttazzo:2020bto} to encourage the experimental collaboration to add a slowly varying background component in the fitting procedure. 
A novel observation we point out in this work is that there is no conceptual obstruction to use data-taking cycles with generic time intervals (i.e. with a duration different than 1 year).

This paper is organized as follows: 
in Section \ref{sec:data} we briefly describe the experimental data considered in our study and we motivate the possibility to use generic time intervals. 
In Section \ref{sec:bayesapproach} we  explain the details of our analysis, performed using the Bayesian approach. 
In Section~\ref{sec:results} we show our results and in particular we give a quantitative comparison between the various models in terms of Bayesian factors, likelihood ratios and Ockham's factors. Finally, conclusions are given in Section \ref{sec:concl}.

\section{The DAMA and DAMA/LIBRA data}
\label{sec:data}
The DAMA and DAMA/LIBRA experiments use ultra-radiopure NaI(TI) scintillating crystals as active target, coupled with photomultipliers to measure the amount of deposited energy.  
The single-hit scintillation events rate is used to look for a possible signal of DM interactions with matter over the large backgrounds due to natural radioactivity of the detector and surrounding environment.
The event rate, expressed in in cpd/kg/keVee (where 'ee' stands for electron equivalent), has only been published~\cite{Bernabei:2000qi, 0804.2741, 1805.10486} after the subtraction of its time-average over each cycle (of roughly one-year duration) following the procedure outlined in the previous section.
Each cycle starts every year roughly around the beginning of September.

The residual rates are available in different energy windows: for the DAMA/NaI and the  DAMA/LIBRA Phase I experiments, the energy windows are (2-4), (2-5), and (2-6) keVee; for the DAMA/LIBRA Phase II the energy windows are (1-3), (1-6), and (2-6) keVee. The (2-6) keVee energy  window is the one with the smallest uncertainties, and  it is common to the three phases.
For this reason, to give a quantitative and fair comparison of the different hypotheses in the three stages we decided to study only the residuals in the (2-6) KeVee energy window. 

In principle, our analysis could be easily extended to other energy bins, although we believe our conclusions would not change significantly.

\subsection{A possible bias in the signal subtraction}
\label{subsec:signal}
The algorithm used by the DAMA collaboration to extract the  time-dependent residual rate from the data has a two-fold objective: account for any constant or slowly varying component while keeping the sinusoidal time structure of an hypothetical DM signal. These requirements put constraints on the optimal time intervals of the data-taking cycle.

The variance of the sinusoidal signal  is given by the quantity $(\alpha - \beta^2)$, with the time averages $\alpha = \langle\cos^2 2\pi/T(t-t_0)\rangle$ and $\beta = \langle\cos2\pi/T(t-t_0)\rangle$. For $\Delta = T$, the quantity $(\alpha - \beta^2)$ is $0.5$ (i.e. $\beta=0$). In ref.~\cite{1805.10486} the value of 0.5 is taken as a figure of merit to asses whether the detectors were operational evenly throughout the period of the modulation of one year.
To the best of our understanding a value of $(\alpha - \beta^2)$ much different than 0.5 implies that either the average of the signal is not vanishing ($\beta \neq 0$), or the data-taking period is not close to one year, or both.

The DAMA collaboration used the argument that $(\alpha - \beta^2)\neq 0.5$ to remove the first cycle of data-taking in table~1 of ref.~\cite{1805.10486} from the  the analysis.

\begin{figure}[!ht]
    \centering
    \includegraphics[width=.50\textwidth]{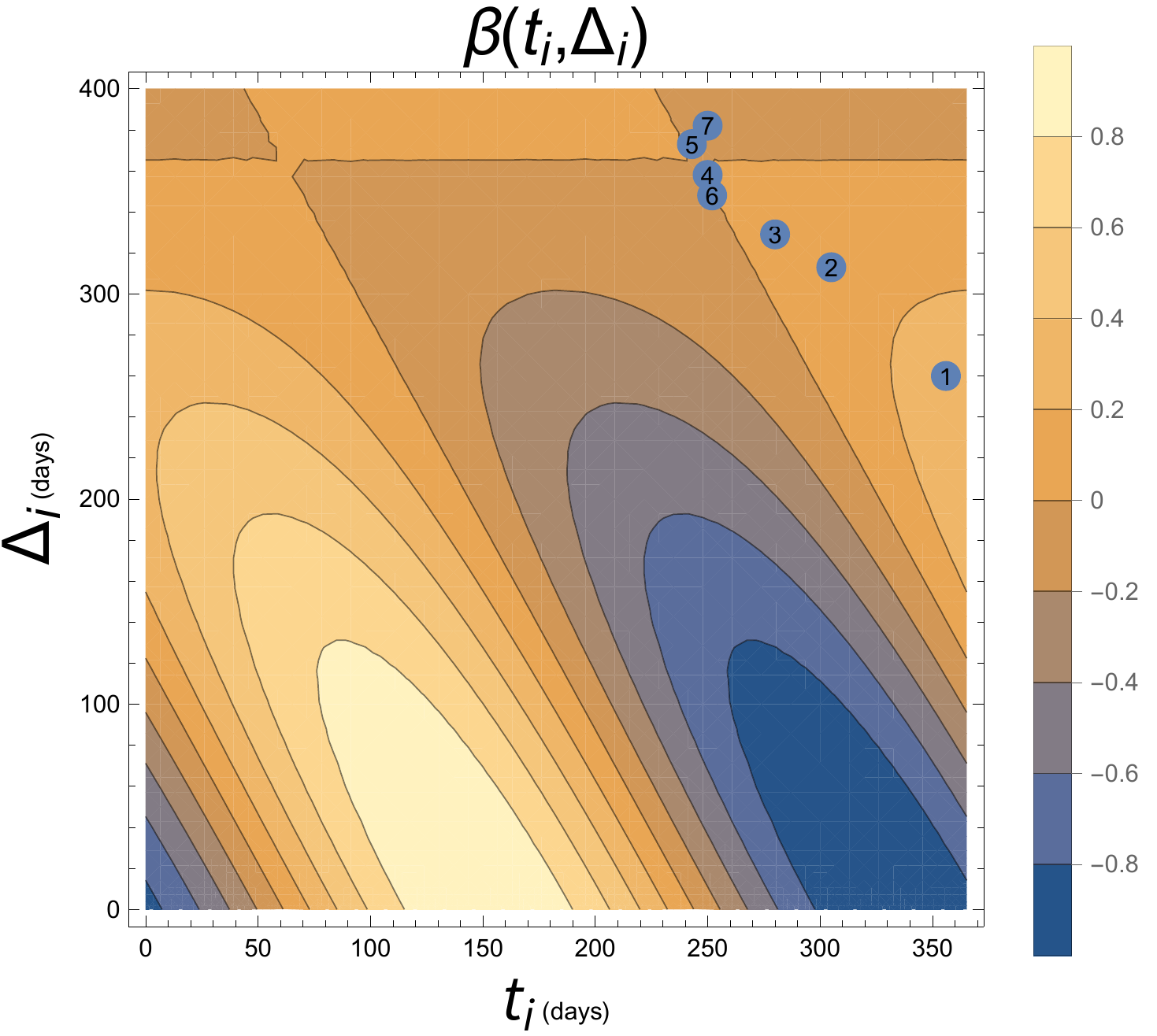}
    \includegraphics[width=.44\textwidth]{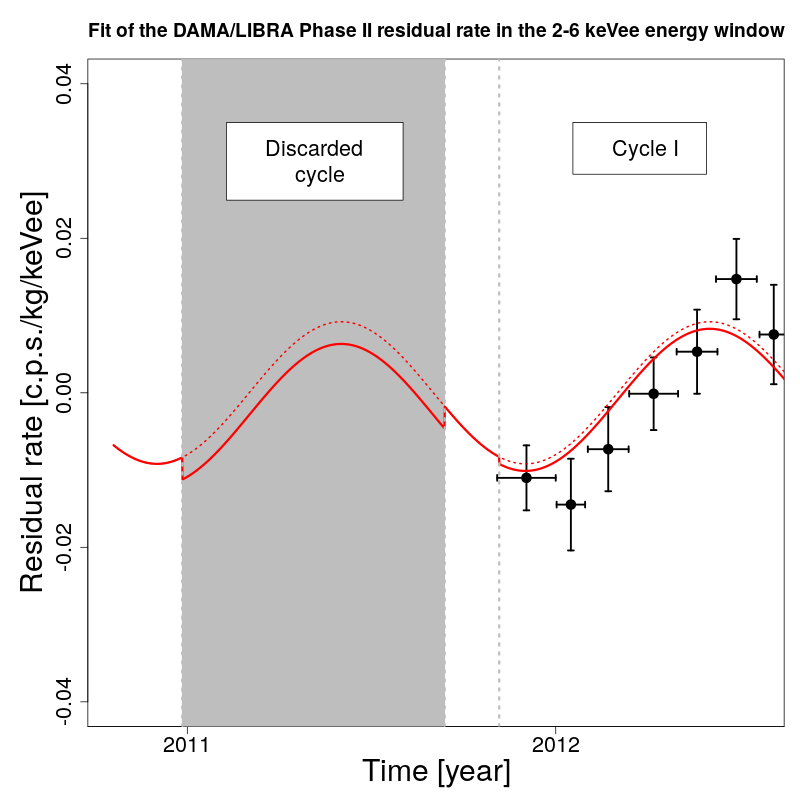}
    \caption{\label{fig:marco} \em {\bf Left}: Contour plot of the function $\beta$ given in eq.~(\ref{eq:bias}); the points represent the data-taking cycles given in table~1 of ref.~\cite{1805.10486}. In particular, the point number 1 corresponds to the first and discarded data-taking cycle. {\bf Right}: Fit of a sinusoidal model (red) to the DAMA/LIBRA Phase II residual rate in the (2-6) keVee energy window obtained taking into account the subtraction-bias effect. The dotted red line is the corresponding sinusoidal modulation without the subtraction. The grey are is the region discarded as explained in ref.~\cite{1805.10486}: since, assuming the sinusoidal signal, the effect is perfectly quantifiable, in principle also this region could be used for the residuals analysis.}
\end{figure}

The procedure to extract the residual rate might induce a subtraction of the signal if the time interval is not chosen carefully. In particular for a data-taking starting in $t_i$ and extending to $t_i+\Delta_i$, the average of the signal is a function of $t_i$ and $\Delta_i$:
\begin{equation}
    \beta(t_i, \Delta_i) = \frac{1}{\Delta_i}\int_{t_i}^{t_i+\Delta_i} \cos\left(\frac{2\pi}{T}(t'-t_0)\right) dt'.
    \label{eq:bias}
\end{equation}
The contour plot of $\beta(t_i, \Delta_i)$ as a function of $t_i$ and $\Delta_i$ is given in fig.~\ref{fig:marco}. Independently of $t_i$, if $\Delta_i$ is equal to $T$, as expected, $\beta(t_i, \Delta_i=T)=0$. This is the approximate condition used by the DAMA collaboration to include a data-taking cycles in the modulation analysis. However, as it is shown in fig.~\ref{fig:marco}, there are other combination of $t_i$ and $\Delta_i$ that result in $\beta(t_i, \Delta_i)=0$; of particular relevance there is the one with $t_i$  and $t_i+\Delta_i$ chosen symmetrically with respect to the time when the cosine is zero. Figure~\ref{fig:marco} also reports the values of $\beta(t_i, \Delta_i)$ for the 7 data-taking cycles reported in ref.~\cite{1805.10486}; we note that for the first and the second cycle the relative contribution of the signal average is of the order of 10\%. In particular, we computed the effect of such a subtraction for the first and discarded cycle and represented it graphically in the plot on the right side of fig.~\ref{fig:marco}. The plot shows the expected behaviour in the discarded region (shaded area): the dotted line corresponds to a sinusoidal model with no subtraction (as generally used), and the continuous line is the corrected sinusoidal model. 
The difference between the dotted and continuous lines corresponds to the value of $\beta$ in the two considered cycles.
Although this effect is generally small in the cycles selected by the DAMA collaboration, we suggest to account for it in the fitting procedure and use
the entire dataset without  being forced to drop any data-taking cycle, even if completely asymmetrical. It would be sufficient to compute the integral in eq.~(\ref{eq:bias}) and include it in eq.~(\ref{eq:St}) potentially as a function of $A$, $T$, and $t_0$.

Finally, we stress that if the background is not constant as a function of time it will have an impact on the value of the average of the rate in the relative data-taking cycle.
In particular
the linear term of expansion of the background as a function of time will produce a sawtooth-shape in the residual rate.

\section{Bayesian approach to the analysis of the DAMA residuals}
\label{sec:bayesapproach}
 The possibility to associate a probability value to any uncertain quantity is at the basis of the Bayesian approach to data analysis.
 This may apply to the outcome of a measurement before it is carried out, to a parameter of interest within a given model, as well as to the model itself. Models and outcomes of measurements are then related by the rules of probability theory. We can compute, at least in principle, the probability of any specific propositions given some state of information. This is particularly useful in the context of model comparison: given two or more models, we can compute the probability that the actual experimental observations are in support of each of them, and then rank the models according to decreasing probability.     

In the following, we summarize how models can be compared in the Bayesian approach, while in Appendix~\ref{app:bayesian} we give all the details. 
Let's suppose to have a dataset $D =\{x_i\}$ composed of $n$ measurements $x_i$, a set of models or hypotheses $H_i$ where each hypothesis could in general depend on a vector of parameters $\vec{\theta_i}$. Within a model, we assume to know the prior probability $\pi(H_i)$, the likelihood function $\mathcal{L}(H_i, \vec{\theta_i}; D)$, and the prior probability density function (pdf) of all the model's parameters $\pi(\vec{\theta_i} | H_i)$.

The posterior odds ratio is given as:
\begin{equation}
  \frac{p \left( H_A|D \right)}{p\left(H_B|D\right)} =
  BF_{AB} \times \frac{\pi(H_A)}{\pi(H_B)},
  \,\,\,\,\,\, {\rm with}\,\,\,\,\,\,  
  BF_{AB} = \frac{\mathcal{L}(H_A; D)}{\mathcal{L}(H_B; D)};
\end{equation}
where $\mathcal{L}(H_A; D)$ is the so called marginal likelihood defined as:
\begin{equation}
 \mathcal{L}(H_A; D)
    = \int \mathcal{L}(\vec{\theta}_A, H_A; D)
    \pi(\vec{\theta}_A|H_A)\, d\vec{\theta}_A.
\label{eq:lik_integral}
\end{equation}  
The Bayes factor $BF_{AB}$ encapsulates all the new information associated to the measurements, although, for parametric models, it could critically depend on the choice of the priors $\pi(\vec{\theta}_i|H_i)$.
In any real situation, the integral of eq.~(\ref{eq:lik_integral}) is not solvable analytically and can only be estimated numerically. Nevertheless, in the case where the information content of the measurements is such that the priors can be considered sufficiently vague in the range of the parameters space where the likelihood is sizable, we can consider the prior as a constant and use the Laplace approximation to estimate the integral. 
This is a very useful approximation to get sense of the different contributions to the Bayes factor because it can be factorised as: 
\begin{equation}
    BF_{AB} = LR_{AB} \times OF_{AB}.
\end{equation}
where we defined $(LR_{AB})$\footnote{
Note that for normal likelihoods the $LR_{AB}$ can be expressed in terms of $\chi^2_{A(B)} =\sum_i\left[ \frac{(x_i - \mu_{A(B)}(x_i))^2}{\sigma_i^2}\right]$ as: $LR_{AB} = \exp{\left[- \frac{\chi^2_A - \chi^2_B}{2}\right]}$,
 where $\mu_{A(B)}(x_i)$ represents the true value of the measurement $x_i$ in the model $A(B)$. This means that the $LR_{AB}$ of the two models evaluated at their best fit values is simply given by $\exp{[-\Delta\hat{\chi}^2_{AB}/2}]$. In a frequentistic approach to model comparison, the $\Delta\hat{\chi}^2_{AB}$ is often taken as test-statistic and its probability distribution used to compute p-values. For nested models, where the more general model contains the simpler plus $k$ additional parameters, the Wilks' theorem says that $pdf(\Delta\hat{\chi}^2_{AB})$ is itself a $\chi^2$ distribution with $k$ degrees of freedom. For non-nested models one would have to estimate $pdf(\Delta\hat{\chi}^2_{AB})$, possibly by sampling it with pseudo-experiments.} as the ratio of the likelihoods  evaluated at their maximum,
and $OF_{AB}$, the so called Ockham's factor, estimated by working out the integral in eq.~(\ref{eq:lik_integral}) as shown in more details in Appendix~\ref{app:bayesian}. 
In this factorization $LR_{AB}$ depends solely on the measurements, while $OF_{AB}$ also on the priors.
The Ockham's factor is a measure of the complexity of the two models that depends on the number and the {\em a-priori} variability of the model's parameters.
The message behind this factorization is that even though a model with more parameters can be more flexible and thus better fit the data producing an higher likelihood, there is a price to pay for having a more complex model. In other words: even if the likelihood ratio pushes towards a more complex model which often better fits the data, on the other hand the Ockham's factor penalises it. 

Finally, two other useful criteria to quantify the ability of a model to describe the observations are the Bayesian Information Criterion (BIC)~\cite{Akaike:1974} and the Akaike Information Criterion (AIC)~\cite{Schwarz:1978}, a review of which is given for example in ref.\cite{Liddle:2007fy}. 
Given a dataset of size $n$, and a  model $A$, with a number $k_A$ of parameters ${\vec{\theta}}_A$ whose value that maximise the likelihood is $\hat{\vec{\theta}}_A$,
the BIC is defined as:
\begin{equation}
    BIC_A = k_A\,\ln{(n)}- 2 \ln{\left(\mathcal{L}(\hat{\vec{\theta}}_A, H_A; D)\right)};
\end{equation}
and the AIC is defined as:
\begin{equation}
    AIC_A =  2\,k_A - 2 \ln{\left(\mathcal{L}(\hat{\vec{\theta}}_A, H_A; D)\right)};
\end{equation}
According to these criteria, the smaller  the value of the BIC (AIC) is, the better the description of the observations. 

When comparing two models,  the $\Delta{}BIC_{AB}$ and $\Delta{}AIC_{AB}$ can be defined as:
\begin{eqnarray}
    \Delta{}BIC_{AB} &= BIC_A - BIC_B =&  \ln{(n)} \left(k_A-k_B\right)  - 2  \ln{\left(LR_{AB}\right)}. \\
    \Delta{}AIC_{AB} &= AIC_A - AIC_B =& 2 \left(k_A-k_B\right) - 2  \ln{\left(LR_{AB}\right)}.
\end{eqnarray}
The previous equations show how the $\Delta{}BIC_{AB}$ and $\Delta{}AIC_{AB}$ are closely related to the $LR_{AB}$, plus a penalty term that penalises the model which is more complex.

For the interpretation of the odds ratio or Bayes factor we refer to the criterion based on Jeffreys scale \cite{ref:Jeff}: a value of $> 10$ represents strong evidence in favor of model A, and a value of $> 100$ represents decisive evidence. Similar evaluation can be done based on the $\Delta{}BIC$ and the $\Delta{}AIC$.

\subsection{Models considered in the analysis}
\label{subsec:models}

We decided to consider, in our analysis, three different models:
\begin{itemize}
    \item COS model: this is the pure-cosine model, where the only free parameter is the amplitude $A$ of the modulation:
    \begin{equation}
        S_{COS} (t) = A \cos{\left(\frac{2 \pi}{T} \left(t-t_0\right)\right)},
    \label{eq:cos_model_residuals}
    \end{equation}
    where the period $T$ is assumed to be fixed to 1 year and $t_0 = 152.5\:d$ such that the peak of the modulation is on the $2^{\text{nd}}$ June.
    \item SAW model: this is the pure-sawtooth model, where the only free parameter is the slope $B$ of the linearly varying sawtooth:
    \begin{equation}
        S_{SAW} (t) = B \left(t - t_i\right) \quad\text{with}\quad
        t_i-\frac{\Delta_i}{2} < t < t_i + \frac{\Delta_i}{2},
    \label{eq:saw_model_residuals}
    \end{equation}
    where $\Delta_i$ is the width of the time window to which $t$ belongs, and $t_i$ is the center of this time window.
    \item C+S model: this is the cosine plus sawtooth model, where this time the free parameters to be considered are two, $A$ and $B$:
    \begin{equation}
        S_{C+S} (t) =  S_{COS}(t) + S_{SAW}(t),
    \end{equation}
    where $S_{COS}$ and $S_{SAW}$ are defined in the eq.~(\ref{eq:cos_model_residuals}) and (\ref{eq:saw_model_residuals}), respectively.
\end{itemize}

What we would like to stress is that the models have been chosen in such a way that the COS and the SAW model have the same number of parameter, in order to minimise the impact of the Ockham's factor (and thus of the priors of the parameters) in the final Bayes factor. Indeed, we expect that, if the C+S model doesn't give a strong improvement in the likelihoods with respect to the other two models, the Ockham's factor will drive the Bayes factor in favour of the simpler models.

\subsection{Implementation of the models using {\tt JAGS} and {\tt rjags}}\label{subsec:implementation}
The computation of the posterior probabilities, even for models relatively simple as those described in the previous section, is often only possible by Monte Carlo integration. The most common way to solve problems of this kind is by sampling the unnormalised posterior distribution by a Markov Chain Monte Carlo (MCMC).
For our study we used the general analysis framework {\tt R} \cite{ref:R} and the MCMC algorithm called {\it Gibbs Sampler} as implemented in  {\tt JAGS} \cite{ref:jags} and interfaced with {\tt R} in the package {\tt rjags} \cite{ref:rjags}. The details of the implementation are given in Appendix \ref{appendix:JAGS}.

We assume, as it is done by the DAMA collaboration \cite{DAMA}, that the measurements at different times are independent and follow a normal likelihood in each time bin $t_i$ with known standard deviation $\sigma_i$ given by the experimental uncertainty. The total likelihood is then given by:
\begin{equation}
    \mathcal{L}(\{\mu_i\}, \{\sigma_i\};\{D_i\} ) = \prod_{i=i}^{n} \frac{1}{\sqrt{2\pi}\sigma_i}\exp\left [-
    \frac{(y_i - \mu_i)^2}{2\sigma_i^2}\right],
\end{equation}
where the true parameters $\{\mu_i\}$ are given as a function of $\{t_i\}$ by implementing one of the three models outlined above, for example for the COS model:
\begin{equation}
    \mu_i = A\, \cos\left ( \frac{2\pi}{T} (t_i - t_0) \right)
\end{equation}
with the only unknown parameter represented by the amplitude $A$.
For each unknown parameter in the model we have to give a prior distribution probability. In the following we used flat priors for all parameters.

The implementation in {\tt JAGS} of the model outlined above and the discussion on the prior's choice is described in Appendix \ref{appendix:JAGS}. The Monte Carlo simulation gives the unnormalised posterior probability of the parameters of interest sampled using the Gibbs algorithm. The results reported in this work are obtained with a single Markov chain with $5\cdot10^5$ steps.
Finally, to estimate the marginal likelihood and the Bayes factor we used the {\tt bridgesampling} package \cite{ref:bridgesampling}. This package uses the same Markov chain used to sample the posterior probability for the integration of the marginal likelihood. In the following, to get a feeling of the contribution of the priors we also report the value of the maximum of the log-likelihood (best fit $\chi^2$);  when comparing two models, we computed analytically the likelihood ratio, and the Ockham's factor as described in Appendix~\ref{app:bayesian} using eq.~(\ref{eq:bayesfactor_example}).

\subsection{Performance of model comparison criteria with simulated data}
We used a toy simulation to qualify the performance of the model selection criteria outlined above on the specific problem discussed in this work.
We generated a few cycles of pseudo-data with the same time structure of those of DAMA, and residual rates sampled from independent normal pdfs with true values given by a mixture of COS and SAW models and known $\sigma$. We choose a value of $A$ for the COS model, and $B$ for the SAW model similar to the amplitude of the DAMA/LIBRA phase II residuals, and a $\sigma$ comparable with their uncertainties.  
We fitted these pseudo-datasets with the three hypotheses: COS, SAW and C+S, and checked the sensitivity of the model comparison criteria to indicate the correct model used in the generation. The source code for the simulated analysis is publicly available on {\tt GitHub}~\cite{ref:git}.

\begin{figure}[!ht]
    \centering
    \includegraphics[width=.48\textwidth]{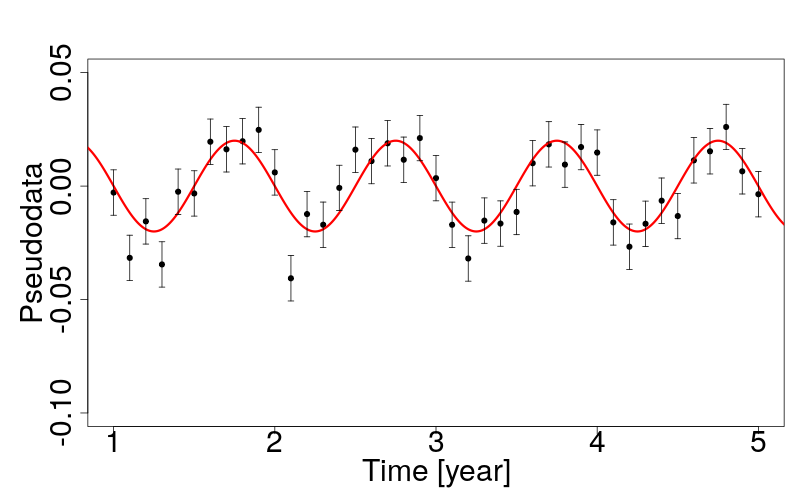}
     \includegraphics[width=.48\textwidth]{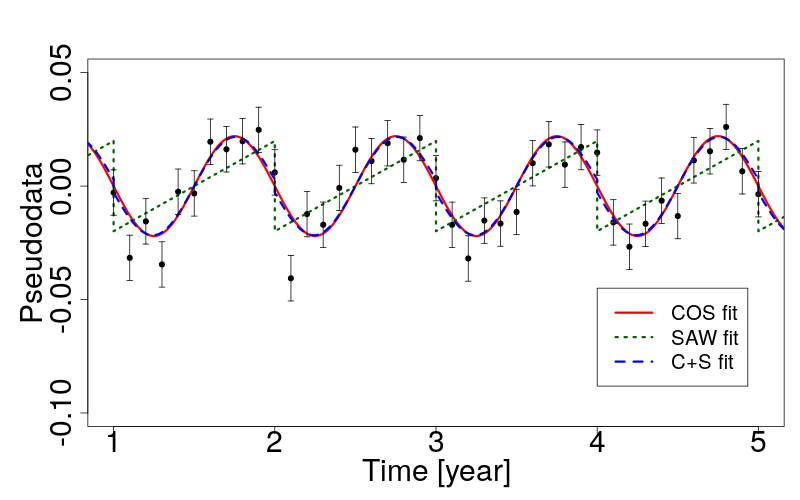}
    \includegraphics[width=.48\textwidth]{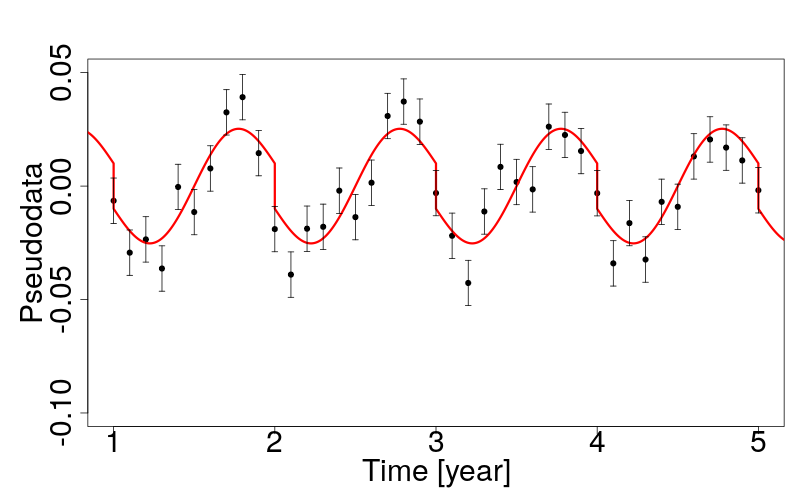}
    \includegraphics[width=.48\textwidth]{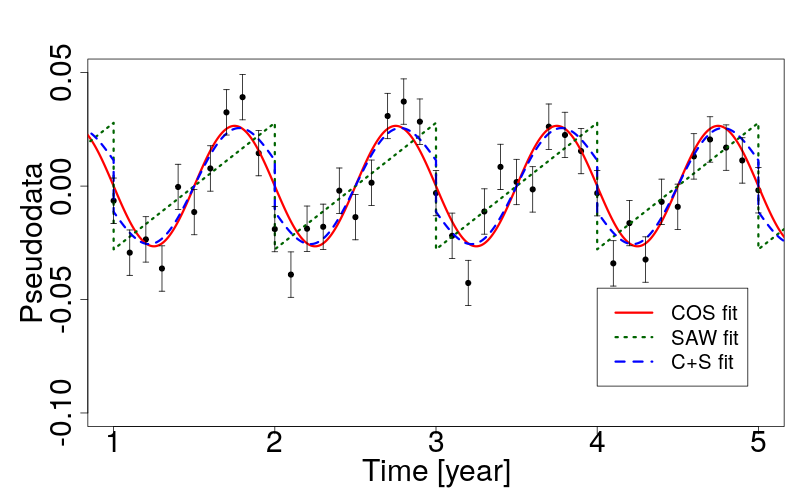}
    \includegraphics[width=.48\textwidth]{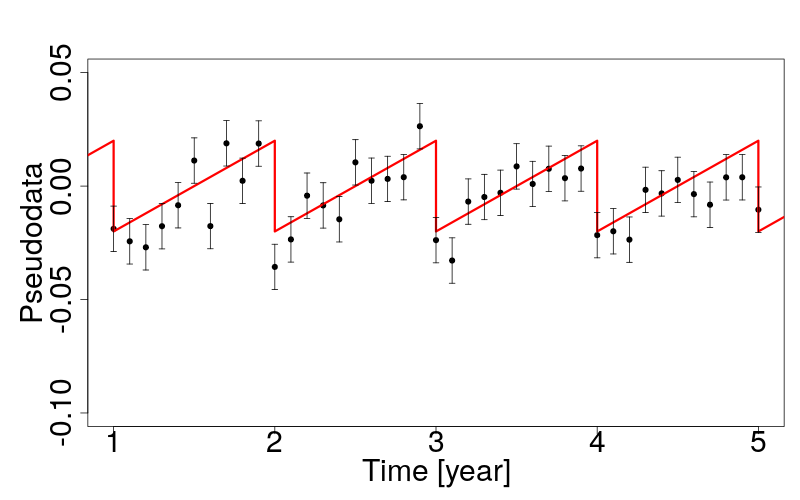}
     \includegraphics[width=.48\textwidth]{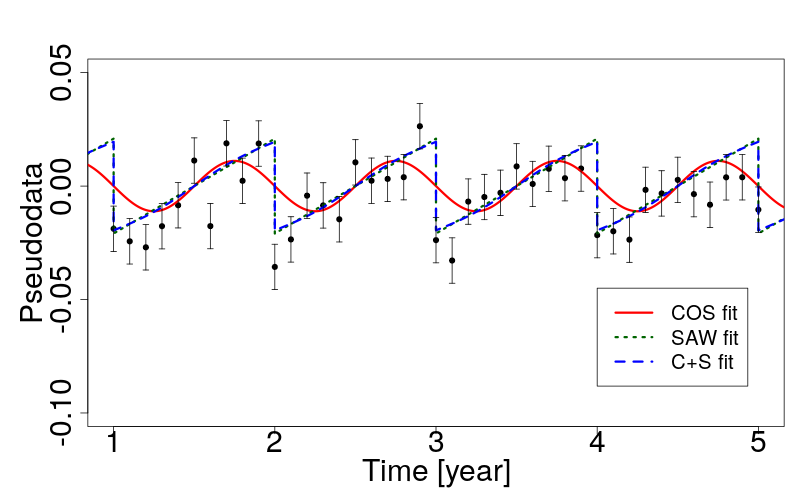}    
    \caption{\label{fig:toy} \em {\bf Left:} Simulated data (points) and true underlying model (red line)  for the COS, C+S, and SAW models respectively. {\bf Right:} best fit  for the COS (red solid line), SAW (green dotted line), and C+S (blue dashed line) models respectively.}
\end{figure}

Figure~\ref{fig:toy} shows the pseudo-datasets with the underlying true model on the left-hand side and the best fit for each of the three models on the right-hand side.
Table~\ref{tab:toy} reports for the three pseudo-datasets the values of the model comparison criteria. We can see that in all cases the underlying model is correctly selected
for each of the criterion of model comparison. The largest scores occur when we compare COS and SAW models, but since C+S is an extension of both COS and SAW, in some cases that include C+S we may have a relatively small BF (of the order of 10), mostly due to OF contribution, namely the difference in complexity between the two compared models.
\begin{table*}[t]
\def\arraystretch{1.1}
\centering
\begin{tabular}{|c||c|c|c|c|}
\hline
COS generated data & BF [dB] & LR [dB] & $\Delta$BIC & $\Delta$AIC \\
\hline
COS vs SAW & $86.4$ & $84.1$ & $-38.7$ & $-38.7$\\
\hline
COS vs C+S & $8.98$ & $-1.26$ &  $-3.13$ & $-1.42$\\
\hline
SAW vs C+S & $-77.4$ & $-85.3$ &  $35.6$ & $37.3$\\
\hline\hline
C+S generated data & BF [dB] & LR [dB] & $\Delta$BIC & $\Delta$AIC \\
\hline
COS vs SAW & $62.9$  & $60.6$  & $-27.9$ & $-27.9$\\
\hline
COS vs C+S & $-9.34$ & $-19.6$ &  $5.30$ & $7.01$\\
\hline
SAW vs C+S & $-72.2$ & $-80.2$ &  $33.2$ & $34.9$\\
\hline\hline
SAW generated data & BF [dB] & LR [dB] & $\Delta$BIC & $\Delta$AIC \\
\hline
COS vs SAW & $-82.6$ & $-85.0$ & $39.1$ & $39.1$\\
\hline
COS vs C+S & $-68.6$ & $-82.0$ & $34.0$ & $35.7$\\
\hline
SAW vs C+S & $14.0$ & $3.00$ & $-5.10$ & $-3.38$\\
\hline
\end{tabular}
\caption{\label{tab:toy} \em Comparison between the various models in the three pseudo-datasets in terms of Bayes factor (BF), likelihood ratio (LR), and difference of Bayesian Information Criterion ($\Delta{}$BIC) and Akaike Information Criterion ($\Delta$AIC).}
\end{table*}

\begin{table*}[t]
\def\arraystretch{1.1}
\centering
\begin{tabular}{|c||c|c|c|c|c|}
\hline
DAMA phase & $BF_{C, S}$ [dB] & $LR_{C, S}$ [dB] & $OF_{C, S}$ [dB] & $\Delta{}$BIC & $\Delta{}$AIC\\
\hline
\hline
DAMA/NaI & $-16.7$ & $-18.1$ & $1.4$ & $8.34$ & $8.34$\\
\hline
LIBRA Phase I & $14.0$ & $12.0$ & $2.0$ & $-5.53$ & $-5.53$\\
\hline
LIBRA Phase II & $86.5$ & $84.7$ & $1.8$ & $-39.0$ & $-39.0$\\
\hline
\hline
All & $88.8$ & $64.7$ & $24.1$ & $-39.7$ & $-33.8$\\
\hline
\end{tabular}
\caption{\label{tab:bayestable} \em Bayes factor (BF), likelihood ratio (LR), Ockham's factor (OF), Bayesian Information Criterion (BIC) difference and Akaike Information Criterion (AIC) difference between the COS and the SAW model for the three stages of DAMA experiments. For the first three rows, the $\Delta{}$BIC and the $\Delta{}$AIC values are equal because the two models have the same number of parameters. The last row refers to the comparison between COS and a SAW model with 3 different slopes as detailed in Section \ref{subsec:model123}. }
\end{table*}

\section{Results}
\label{sec:results}

As already said at the beginning of the Section \ref{sec:bayesapproach}, we decided to analyse the residual rate in the (2-6) keVee energy window. 
Since the precision of the measurements and the size of the modulation are different in the various phases, as a first step we analysed the data individually on each of the three datasets, testing only the two hypotheses of pure-cosine (COS) and pure-sawtooth (SAW) contribution. We tested on the most informative dataset also the third model (C+S), and then we performed the analysis on the whole dataset. Finally, we evaluated the performance of the cosine model promoting the period and the phase to free parameters.

\subsection{COS versus SAW models}
\begin{figure}[!ht]
    \centering
    \includegraphics[scale = 0.26]{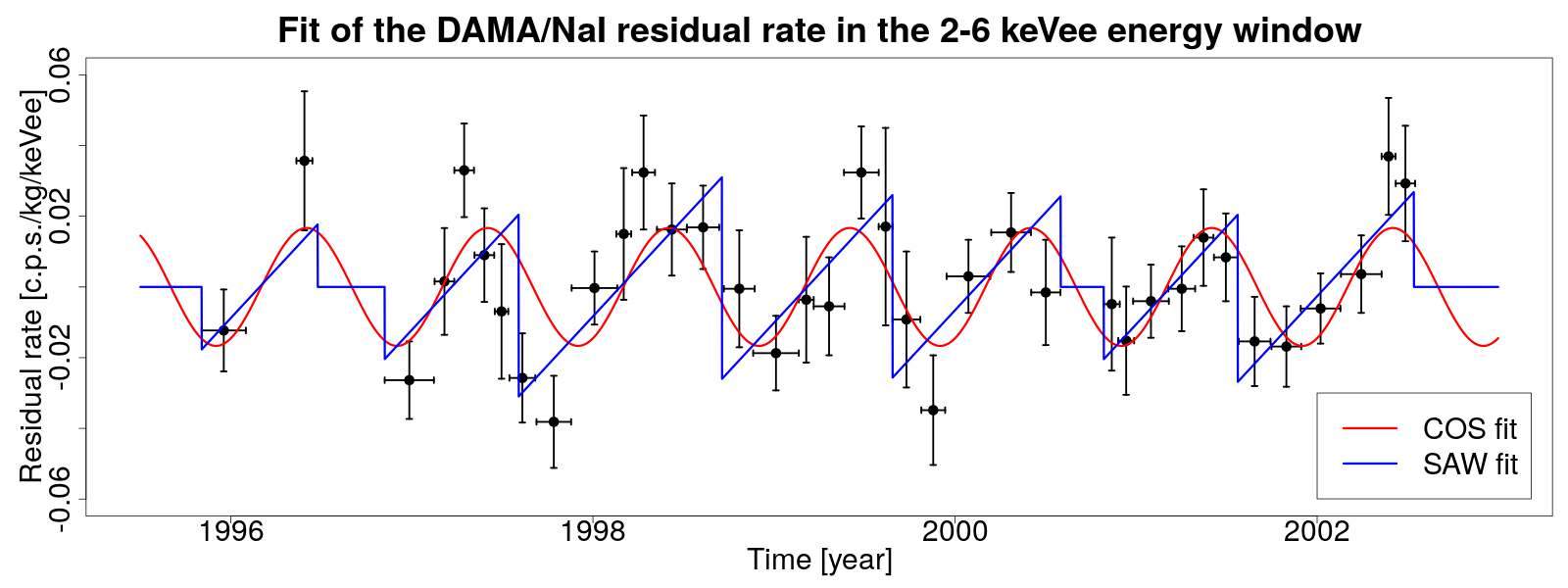}
    \caption{\label{fig:naifit} \em Fit to the DAMA/NaI residual rate in the (2-6) keVee energy window. The regions in which the blue line is constant (and equal to zero) are the regions between two non-contiguous data-taking cycles. }
\end{figure}

\begin{figure}[!ht]
    \centering
    \includegraphics[scale = 0.26]{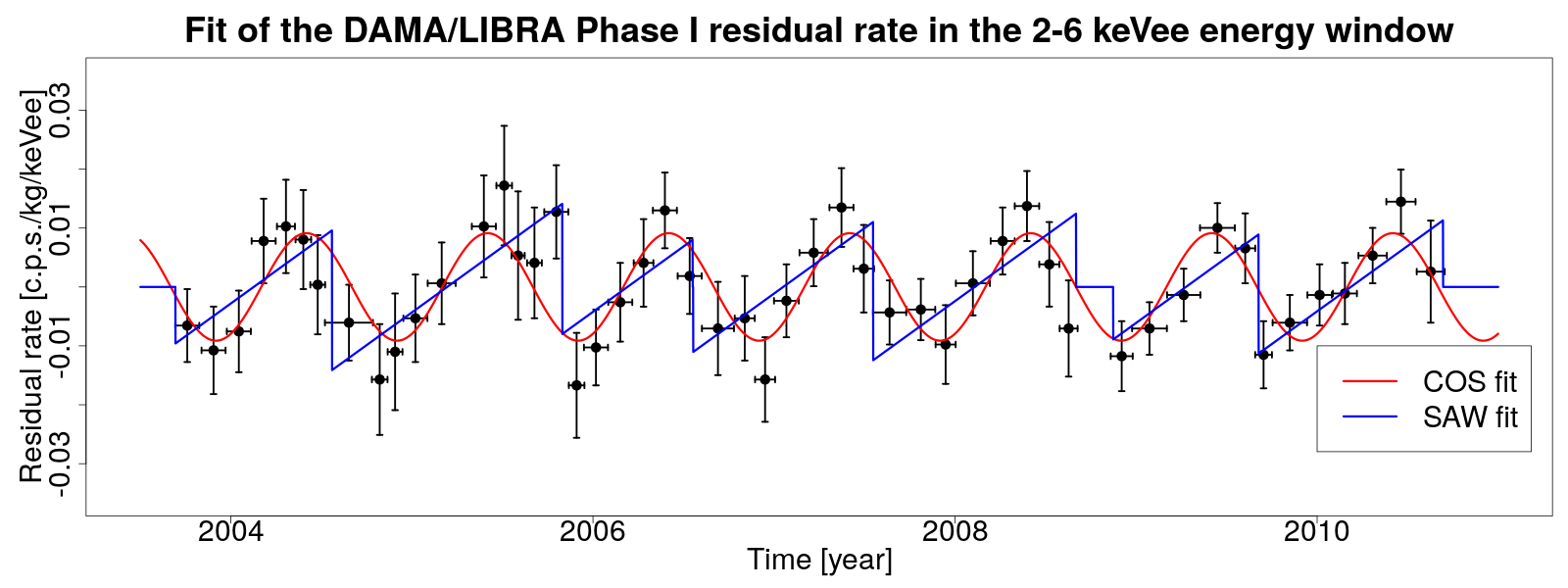}
    \caption{\label{fig:lib1fit} \em Fit to the DAMA/LIBRA Phase I residual rate in the (2-6) keVee energy window. The regions in which the blue line is constant (and equal to zero) are the regions between two non-contiguous data-taking cycles.}
\end{figure}

\begin{figure}[!ht]
    \centering
    \includegraphics[scale = 0.26]{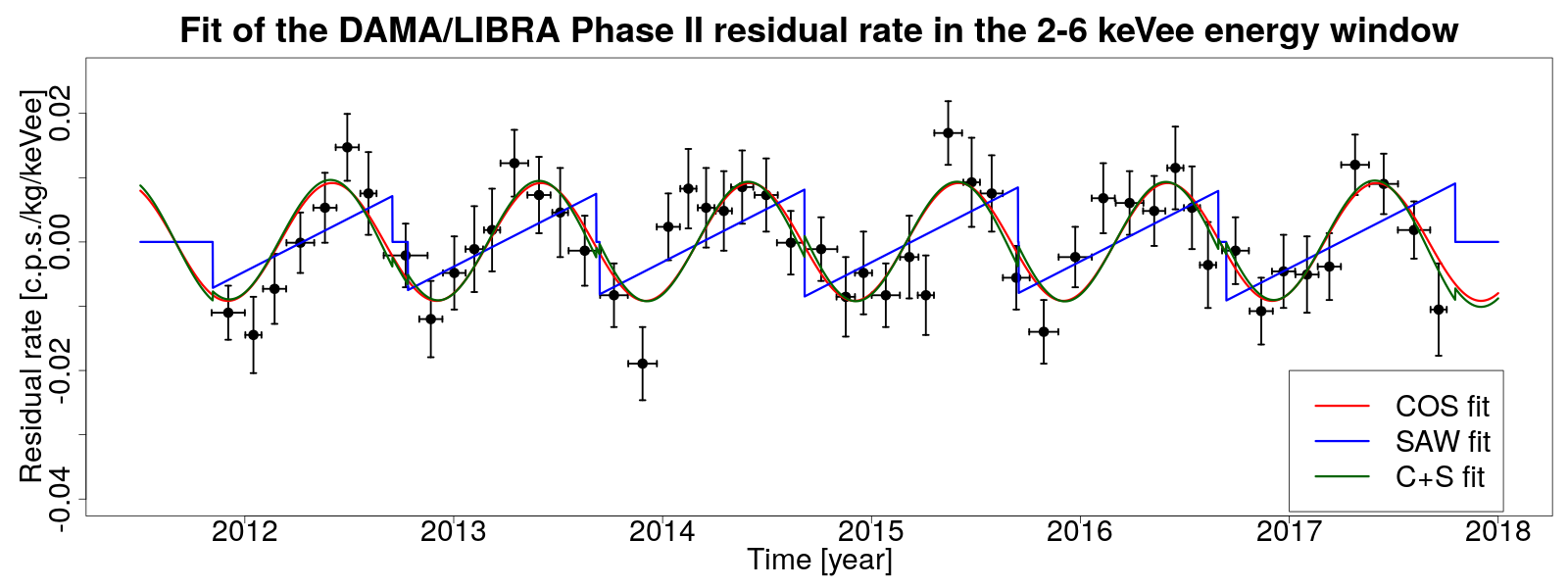}
    \caption{\label{fig:lib2fit} \em Fit to the DAMA/LIBRA Phase II residual rate in the (2-6) keVee energy window. The regions in which the blue line is constant (and equal to zero) are the regions between two non-contiguous data-taking cycles.}
\end{figure}

In table~\ref{tab:bayestable} we show the results of the comparison between the COS and the SAW models in terms of Bayes factor, likelihood ratio, Ockham's factor, Bayesian Information Criterion difference and Akaike Information Criterion difference in every experimental stage. Since some of these quantities can eventually be very large or very small, BF, LR and OF are given in decibels\footnote{Given a quantity $x$, the decibel $d$ is defined as $d=10\log_{10}(x)$. }. This table also shows how the contribution of the Ockham's factor, which is the component of the Bayes factor critically dependent on the choice of the priors of the parameters, is marginal. As explicitly shown in Appendix \ref{appendix:JAGS}, for each of the three datasets we chose a flat prior on $B$ as well as on $A$. However, for all the three datasets we tested different possible priors (uniform, normal and gamma distributions), but in all cases the Ockham's factor contribution is always under control, namely it changes of at the most 4-5 units in decibels with respect to our final choice of the prior. The results for the different stages can be summarized as follows:
\begin{itemize}
    \item For the DAMA/NaI dataset, the Bayes factor, which is $10^{-1.67}$ in normal units, disfavours the pure-cosine model in favour of the pure-sawtooth one. However this is the less informative dataset, namely it is the one with relative uncertainties on each point larger than in the other phases. The fit is shown in fig.~\ref{fig:naifit}.
    \item For the DAMA/LIBRA Phase I dataset, the Bayes factor is of the order of $10^{1.40}$, and this means that already at this stage, where the relative uncertainties are smaller than in the previous case, the cosine model starts to win on the pure-sawtooth model.  The fit is shown in fig.~\ref{fig:lib1fit}.
    \item For the DAMA/LIBRA Phase II dataset, the Bayes factor is $10^{8.65}$. This means that in the region in where the data give the maximum of the information, in the sense that we explained before, the pure-cosine model is greatly preferred to the pure-sawtooth model. In fig.~\ref{fig:lib2fit} the result of the fit is presented.
\end{itemize}
The details of the three fits are described in table~\ref{tab:fittable}, in which all the reported values are obtained computing the expected value and the standard deviation on the posterior probability of each parameter.


\begin{table*}[t]
\def\arraystretch{1.1}
\centering
\begin{tabular}{|c||cc||cc|}
\hline
DAMA & \multicolumn{2}{c||}{Fit to COS model} & \multicolumn{2}{c|}{Fit to SAW model}\\ 
phase & $A$ [cpd/kg/keVee] & $\chi^2_{\rm cos}/{\rm dof}$ & $B$ [cpd/kg/keVee/yr] & $\chi^2_{\rm saw}/{\rm dof}$ \\
\hline
\hline
DAMA/NaI & $0.0168\pm 0.0029$ & $36.7/36$ & $0.0552\pm 0.0085$   & $28.3/36$ \\
\hline
LIBRA I & $0.0092\pm 0.0013$ & $29.6/49$ & $0.0222\pm 0.0032$   & $35.1/49$ \\
\hline
LIBRA II & $0.0092\pm 0.0011$ & $43.3/51$ & $0.0166\pm 0.0029$   & $82.3/51$ \\
\hline
\end{tabular}
\caption{\label{tab:fittable} \em Results of the fit for the cosine amplitude $A$ of the COS model and the sawtooth coefficient $B$ of the SAW model obtained on the DAMA residuals in the (2-6) keVee energy window during the different phases, together with the corresponding $\chi^2/\text{d.o.f}$.
}
\end{table*}
\subsection{Performance of the COS, SAW, and C+S models}
Since the third dataset is the most informative one, as already shown in fig.~\ref{fig:lib2fit}, we decided to test not only the COS and the SAW model, but also the C+S model, as defined in Section \ref{subsec:models}. The results of the fit are given in table~\ref{tab:cossawfit} together with the results of the comparisons between the various models. The value of $B$ obtained in the C+S model is consistent with zero, $B = (-0.0035 \pm 0.0042)$ cpd/kg/keVee/yr, and the value of $A$ is consistent with the value of $A$ obtained in the COS model. Indeed, looking at the fig.~\ref{fig:lib2fit}, the green (S+C fit) and the red (COS fit) lines are very close: in fact the $\chi^2/dof$ of the C+S fit is very similar to that of the COS fit, or, in other words, the LR of the models is very close to one (it is $10^{-0.14} = 0.72$ in normal units). Therefore the Bayes factor in the COS vs C+S case is driven by the Ockham's factor, and since the number of parameters is different in the two models, the COS model wins against the C+S model. The C+S model is an extension of the COS model but, even if it's more complex, it doesn't improve the goodness of fit, and the price to pay for the greater complexity is reflected in the Bayes factor. Since in this case the value of BF depends critically on the priors, we tried to use different possible priors for the parameters as before, but in all cases the BF favoured the COS model. On the other hand, for the SAW and C+S models, although the Ockham's factor pushes in the direction of the simpler SAW model, the contribution of the LR drives the BF in favour of the C+S model. The final message of this analysis is that the most informative dataset available to us can be better represented by models that contain a dominant cosine component, as the COS and C+S models, with respect to the SAW model. In addition, the sawtooth component of the C+S model, quantified by the $B$ parameter, is compatible with zero.

\begin{table*}[t]
\def\arraystretch{1.1}
\centering
\begin{tabular}{|c||c|c|c|}
\hline
Fit results & $A$ [cpd/kg/keVee] & $B$ [cpd/kg/keVee/yr] & $\chi^2/{\rm dof}$ \\
\hline
\hline
C+S fit & $0.0102\pm 0.0016$ & $-0.0035\pm 0.0042$   & $42.7/50$ \\
\hline
\end{tabular}
\vspace{2mm}
\begin{tabular}{|c||c|c|c|c|c|}
\hline
Model comparison & BF [dB] & LR [dB] & OF [dB] & $\Delta$BIC & $\Delta$AIC \\
\hline
COS vs SAW & $86.5$ & $84.7$ & $1.8$ & $-39.0$ & $-39.0$\\
\hline
COS vs C+S & $11.3$ & $-1.4$ & $18.0$ & $-3.31$ & $-1.36$\\
\hline
SAW vs C+S & $-69.9$ & $-86.2$ & $16.3$ & $35.7$ & $37.7$\\
\hline
\end{tabular}
\caption{\label{tab:cossawfit} \em {\bf Top}: Results of the fit of the cosine amplitude $A$ and the sawtooth coefficient $B$  of the C+S model obtained on the DAMA residuals in the (2-6) keVee energy window during the DAMA/LIBRA Phase II, together with the corresponding $\chi^2/\text{d.o.f}$. {\bf Bottom}: Comparison between the various models in the DAMA/LIBRA Phase II dataset in terms of Bayes factor (BF), likelihood ratio (LR), Ockham's factor (OF) and difference of Bayesian Information Criterion ($\Delta{}$BIC) and Akaike Information Criterion ($\Delta$AIC). For all these four metrics (BF, LR, $\Delta{}$BIC, $\Delta$AIC) the SAW model is largely disfavoured.}
\end{table*}

\subsection{Model comparison on the whole dataset}
\label{subsec:model123}

\begin{figure}
    \centering
    \includegraphics[scale = 0.26]{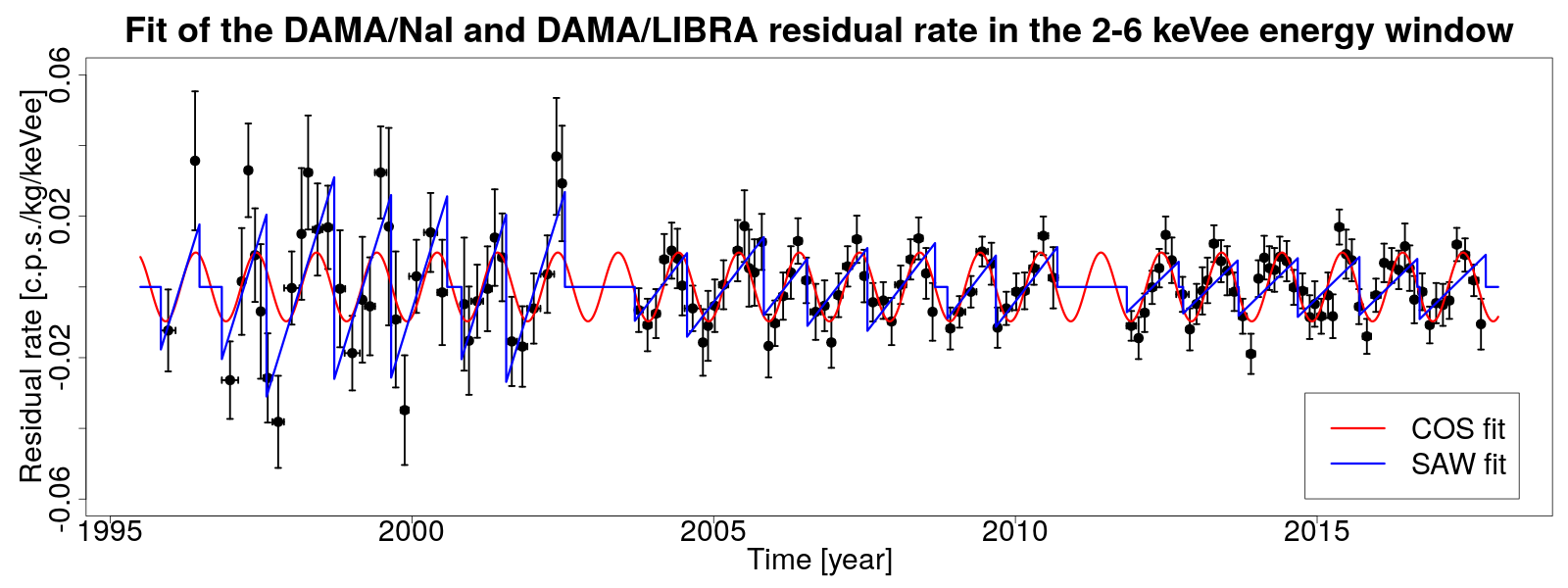}
    \caption{\label{fig:totalfit} \em Fit to the DAMA residual rate in all the three stages in the (2-6) keVee energy window.}
\end{figure}

Since adding the sawtooth component to the cosine modulation in the most informative dataset didn't produce any change with respect to the pure-cosine model, the next step of our analysis was comparing the SAW and COS model on all the available data. In principle in the various phases the size of the sawtooth variation could be different, thus we allowed to have three different $B$ parameters. The results of the global fit is shown in fig.~\ref{fig:totalfit} and the details can be summarized as follows:
\begin{itemize}
    \item For the COS fit:
        \begin{equation}
            A = (0.00973 \pm 0.00078)\: \text{cpd/kg/keVee}
            \qquad 
            \chi^2 / dof = 116.0/138
        \end{equation}
     \item For the SAW fit:
        \begin{equation}
        \begin{cases}
            B_{NaI} &= (0.0553 \pm 0.0085) \: \text{cpd/kg/keVee/yr} \\
            B_{LIBRAI} &= (0.0222 \pm 0.0032) \: \text{cpd/kg/keVee/yr}
            \qquad 
            \chi^2 / dof = 145.8/136 \\
            B_{LIBRAII} &= (0.0166 \pm 0.0028) \: \text{cpd/kg/keVee/yr} \\
        \end{cases}
        \end{equation}
\end{itemize}
As a cross check, for the SAW model in each of the phases we obtain for the parameters $B$ the same results of table~\ref{tab:fittable}. The comparison between COS and SAW model gives
\begin{equation}
\begin{split}
    BF &= 88.8 \: \text{dB}, \\
    LR &= 64.7 \: \text{dB}, \\
    OF &= 24.1 \: \text{dB}, \\
    \Delta{}BIC &= -39.7 \:,\\
    \Delta{}AIC &= -33.8 \:.\\
\end{split}
\end{equation}
In this case the Ockham's factor is still quite marginal with respect to the LR, and the final Bayes factor, which is $10^{8.88}$ in normal units, pushes again strongly towards the COS model.

\subsubsection{Extraction of phase and period of the cosine model}
Finally, we decided to fit all the available data with a pure-cosine model in which the phase and the period of the modulation are not fixed, that is to say that $t_0$ and $T$ defined in eq.~(\ref{eq:cos_model_residuals}) are now treated as parameters. In this case, choosing uniform priors for the two new parameters, we obtain:
\begin{equation}
    \begin{cases}
        A &= (0.00981 \pm 0.00079)\: \text{cpd/kg/keVee} \\
        t_0 &= (0.382 \pm 0.037)\: \text{yr}  \qquad \qquad\qquad\qquad
            \chi^2 / dof = 112.8/136 \\
        T &= (1.0008 \pm 0.0023)\: \text{yr} \\
    \end{cases}
\end{equation}
These results are compatible with a period of 1 year and a $t_0 = 152.5 \: d = 0.418 \: yr$. The comparison between cosine and SAW model this time gives
\begin{equation}
\begin{split}
    BF &= 62.4 \: \text{dB}, \\
    LR &= 71.7 \: \text{dB}, \\
    OF &= -9.3 \: \text{dB}, \\
    \Delta{}BIC &= -33.0 \:,\\
    \Delta{}AIC &= -33.0 \:.\\
\end{split}
\end{equation}
Therefore, again, even if now the SAW model enters the game with the same number of parameters of the cosine model (in fact now the OF slightly favours the SAW model), the BF, which is $10^{6.24}$ in normal units, still pushes strongly towards the cosine model.

\subsection{A suggested model to interpret the DAMA residuals}
\label{subsec:suggestedModel}
All our studies indicate that the time modulation of the DAMA residual rate in the $(2-6)$~keVee energy window cannot be described by an artefact due to the algorithm to subtract a slowly varying background.

Nevertheless, we believe that such an algorithm could potentially have an impact on the extraction of the parameter of interest of the signal.
In particular the definition of the time window used to average the rate can introduce the following problems in the residual rate:
\begin{itemize}
    \item a non zero contribution due to the possible presence of a signal as discussed in Sec.~\ref{subsec:signal};
    \item a sawtooth time modulation due to the presence of a slowly varying background; such a modulation can enhance or reduce the amplitude of a sinusoidal signal, as well as affect its phase.   
\end{itemize}
For this reason we suggest to use a model that takes these effects into account.

Let's consider a data-taking cycle that starts at $t^*$ and extends for a period of time $\Delta$, the true value $\mu_i$ of the observed rate in the time bin $t_i$ is:
\begin{equation}
    \mu_i = A\,\cos{\left( \frac{2\pi}{T}(t_i - t_0)\right)} -
    \frac{A}{\Delta}\int_{t^*}^{t^*+\Delta} \cos\left(\frac{2\pi}{T}(t'-t_0)\right) dt' +
    B\left(t_i - \frac{\Delta}{2}\right)
\end{equation}
where $A$ is the amplitude, $T$ the period, and  $(2\pi\,t_0/T)$ the phase of the sinusoidal signal, while $B$ is the slope of a linearly varying background. For a different time dependence of the background a linear model is however the first order contribution in a time power series.

\begin{figure}[!ht]
    \centering
    \includegraphics[width=.9\textwidth]{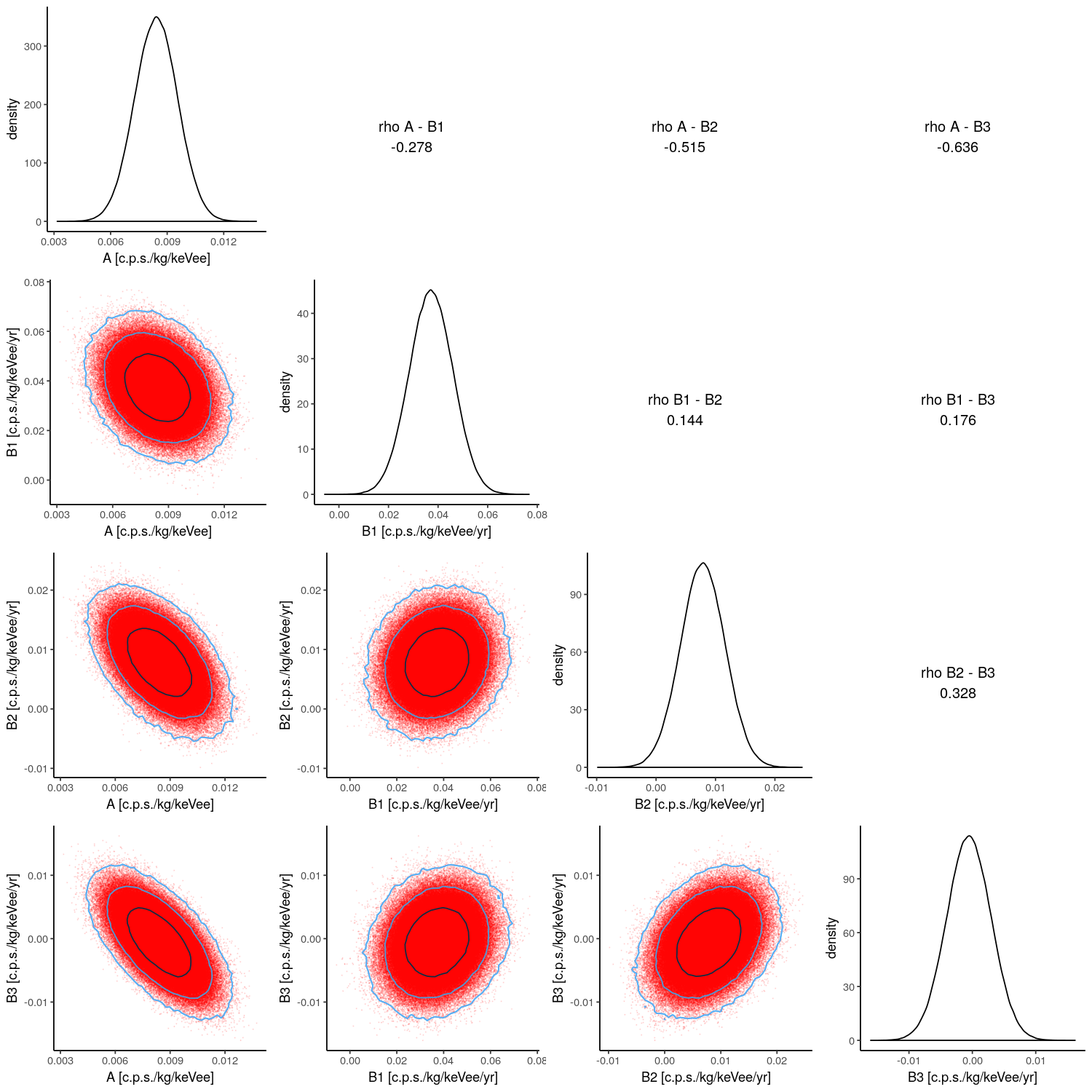}
    \caption{\label{fig:posteriorcomplete4} \em Marginal posterior pdfs for the free parameters of the proposed model (Sec.~\ref{subsec:suggestedModel}) at fixed $T$ and $t_0$ for the fit of the whole (2-6) keVee DAMA dataset.
     The 4 plots on the diagonal of the figure are the uni-dimensional pdfs of each single free parameter obtained by marginalising on the all the others. The 10 bi-dimensional pdfs in the bottom-left corner of the figure give the marginal pdfs of each pair of parameters obtained by marginalising on the other. The plots show also the credible regions at $0.68,0.95,0.997$ probability as black, green, and blue contours respectively. The correlation coefficients are given in the upper-right corner of the figure.}
\end{figure}

\begin{figure}[!ht]
    \centering
    \includegraphics[width=.9\textwidth]{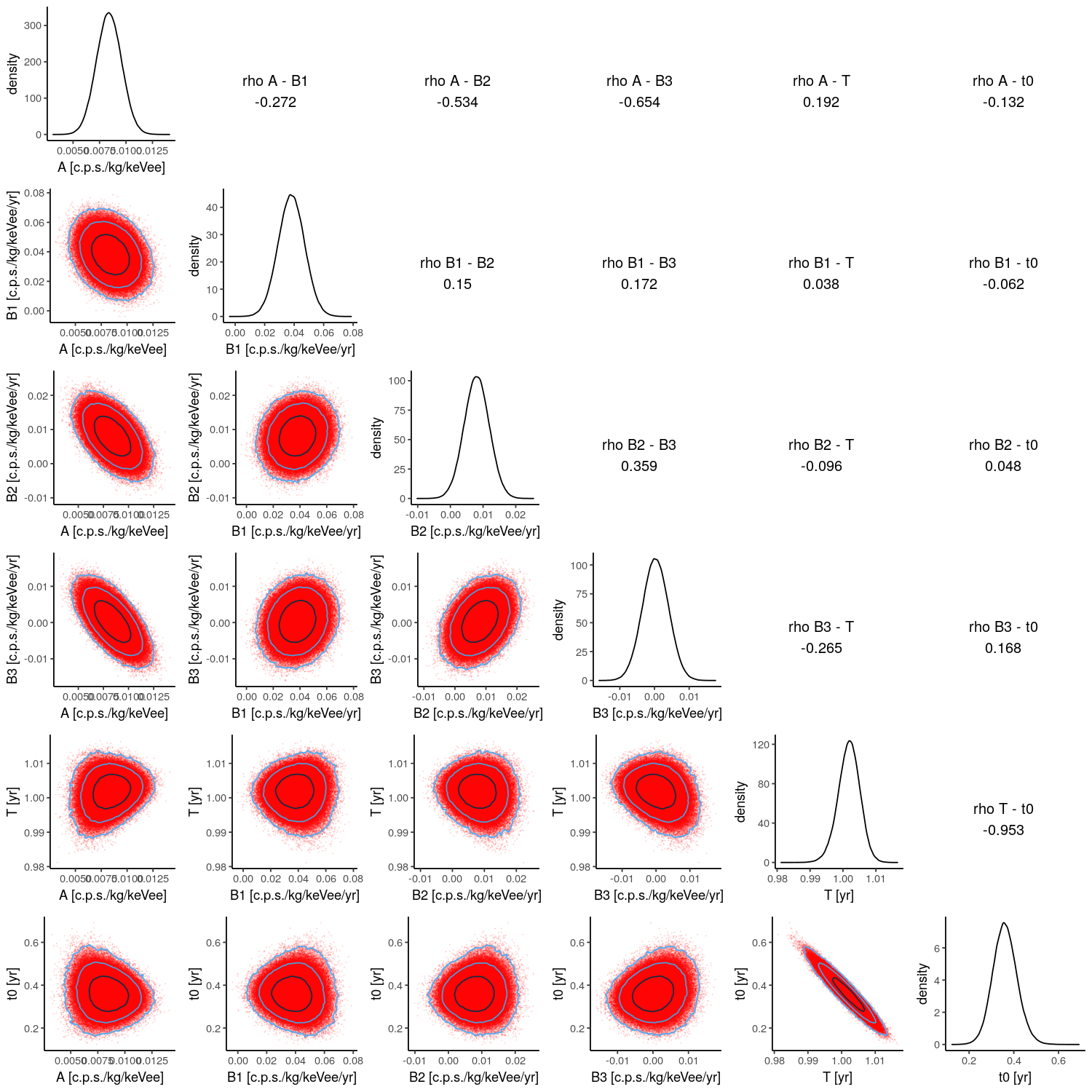}
    \caption{\label{fig:posteriorcomplete} \em Marginal posterior pdfs for the free parameters of the proposed model (Sec.~\ref{subsec:suggestedModel}) for the fit of the whole (2-6) keVee DAMA dataset.
     The 6 plots on the diagonal of the figure are the uni-dimensional pdfs of each single free parameter obtained by marginalising on the all the others. The 15 bi-dimensional pdfs in the bottom-left corner of the figure give the marginal pdfs of each pair of parameters obtained by marginalising on the other. The plots show also the credible regions at $0.68,0.95,0.997$ probability as black, green, and blue contours respectively. The correlation coefficients are given in the upper-right corner of the figure.}
\end{figure}

For the sake of completeness, we deployed such a model to fit the three phases of the experiment in the (2-6)~keVee energy window. Like we did in Section \ref{subsec:model123}, since in the various phases the size of the sawtooth variation could be different, we allowed to have three different $B$ parameters. We performed a first fit (F1) keeping both $T$ and $t_0$ fixed to the values expected for a DM signal, as well as a second one (F2) allowing them to vary.
To better understand the differences with respect to the COS model fit, we decided to show in the fig.~\ref{fig:posteriorcomplete4} and fig.~\ref{fig:posteriorcomplete} the marginal posteriors pdfs of all the fit parameters for F1 and F2, respectively.
\begin{table*}[!ht]
\def\arraystretch{1.1}
\centering
\begin{tabular}{|l||c|c|}
\hline
& F1 & F2 \\
\hline
\hline
A\,[cpd/kg/keVee] & $0.0084 \pm 0.0011$ & $0.0084 \pm 0.0012$ \\
\hline
B1\,[cpd/kg/keVee/yr] & $0.0371 \pm 0.0089$ & $0.0381 \pm 0.0090$  \\
\hline
B2\,[cpd/kg/keVee/yr] & $0.0078 \pm 0.0038$ & $0.0080 \pm 0.0038$  \\
\hline
B3\,[cpd/kg/keVee/yr] & $-0.0006 \pm 0.0035$ & $-0.0003 \pm 0.0038$  \\
\hline
\end{tabular}
\caption{\label{tab:bayestable2} \em Results of the F1 and F2 fit in terms of bet-fit values of the parameters.}
\end{table*}
\\As shown in table~\ref{tab:bayestable2}, except for $B1$ which is positive, the other two are compatible with zero in both cases. The value of $A$ extracted in this way is slightly smaller with respect to the previous cases because of the expected anti-correlation with the $Bi$ parameters. The small correlation between the $Bi$ parameters, assumed to be {\em a-priori} independent, is an induced effect of their common anti-correlation with the $A$ parameter. For the F2 fit, the values of $t_0$ and $T$ are consistent with those expected for a DM signal. Finally, we remark that even though we included into the model the correction for a signal ``double-counting'' as described in Section~\ref{subsec:signal}, this effect is not relevant with the current choice of time intervals for the experimental cycles.


\clearpage
\section{Conclusions}
\label{sec:concl}
In this paper we elaborated on the proposal of ref.~\cite{Buttazzo:2020bto} to interpret the DAMA modulation as a possible effect of a slow varying time dependent background.  We performed a Bayesian analysis of the DAMA residual rate in the (2-6) keVee energy range with the aim of giving a quantitative comparison of the two possible hypotheses: the cosine modulation claimed by the DAMA collaboration and the sawtooth model proposed in ref.~\cite{Buttazzo:2020bto}. The source code used in this work has been made publicly available on {\tt GitHub}~\cite{ref:git}.

For the different models considered we estimated the posterior probabilities, and for each pair of models we computed both the Bayes factor, factorised also in terms of likelihood ratio and Ockham's factor, clearly separating the contributions coming from the data and the prior choice, the Bayesian Information Criterion and the Akaike Information Criterion.

The conclusion of this work can be summarised as follows. With an exception for the DAMA/NaI dataset, which is however the less informative one, in all the comparisons we performed in Section \ref{sec:results} the sawtooth hypothesis is always disfavoured with respect to the cosine modulation hypothesis. This is quantifiable in the Bayes factor which is in all cases between $10^6$ and $10^{8}$. To be more specific, in the case of using the entire dataset and comparing the cosine with just the amplitude as a free parameter in the fit against the sawtooth with three independent and free to vary slope parameters for the three phases of data-taking, we obtain a Bayes factor of $10^{8.88}$ in favour of the cosine model. The effect of the priors choice has been extensively tested and can be reasonably quantified in a contribution at the most of the order $10^{2}$ to the total BF.
Finally, we used the full cosine model with free period and phase and still found a BF of order $10^6$. For this case we obtained a value of the period and phase compatible with those reported by the DAMA collaboration.

Finally, we point out that the background subtraction algorithm used by DAMA can introduce a bias in the parameters of the fitted signal, produced by a residual contribution from the expected signal as well as a contribution from a slowly varying background.
Although the time intervals are chosen in a such a way that this bias is small, we showed that these effects can be safely taken into account in the analysis as shown in Section~\ref{subsec:suggestedModel}. 

\acknowledgments
We thank Dario Buttazzo, Fabio Cappella, and Paolo Panci for useful discussions.
\newpage
\appendix

\section{Bayesian model comparison}
\label{app:bayesian}

In the following we shall suppose to have a dataset $D =\{x_i\}$ composed of $n$ measurements $x_i$, a set of models or hypotheses $H_i$ where each hypothesis could in general depend on a vector of parameters $\vec{\theta_i}$. Within a model,  we assume to know the likelihood function:
\begin{equation}
    \mathcal{L}(H_i, \vec{\theta_i}; D) = 
    p(D | H_i, \vec{\theta_i}, I),
    \label{eq:likelihood_def}
\end{equation}
the expression on the right-hand-side of the previous equation is the probability density function (pdf) to observe some specific data $D$ given the model $H_i$, its parameters $\vec{\theta_i}$, and all the relevant information $I$ before the measurement is carried out. The likelihood, which is numerically identical to $p(D | H_i, \vec{\theta_i}, I)$, has to be intended as a function of the model and its parameters at fixed (observed) data.   
We shall assign a prior probability $\pi(H_i | I)$ to each model, as well as a prior probability $\pi(\vec{\theta_i} | H_i, I)$ to its parameters, both conditioned to the state of information $I$. Priors are intended to encapsulate the probability that the model is valid before data are seen. For the parameters, the priors define their range of variability by setting the probability to find them in any given interval of the allowed space.  In the following, all probabilities are conditional to the state of information $I$ even if, to keep the notation simple, we often don't write it explicitly. 

Suppose we want to compare two different models $H_i$ and $H_j$, which for the moment we assume don't depend on any parameter. 
To compare these models we can compute the ratio of the posterior probabilities:
\begin{equation}
    O_{ij} \equiv \frac{p \left( H_i|D,I \right)}{p\left(H_j|D, I\right)},
\label{eq:odds_def}
\end{equation}
this ratio gives the odds in favour of one model over the other. To express the odds as a function of the likelihoods and priors, we shall apply the Bayes theorem; that is, for example for the numerator of eq.~(\ref{eq:odds_def}), 
\begin{equation}
    p \left(H_i | D, I\right) = \frac{p\left(D|H_i, I\right)
        \pi\left(H_i | I\right)}{\pi \left(D | I\right)}.
\end{equation}
The denominator $\pi(D|I)$ is the prior probability of data and it can be more easily interpreted if expressed in terms of the likelihoods by imposing that $\sum_ip(H_i|D,I)=1$. To do this we need to have a complete class of hypotheses, and this is not often the case. However, if we are only interested in the odds ratio we can neglect this term as it cancels out in the odds.  The  ratio between the probabilities of the two models becomes:
\begin{equation}
    O_{ij} \equiv \frac{p \left( D|H_i,I \right)}{p\left(D|H_j, I\right)}
        \times \frac{\pi\left(H_i|I\right)}{\pi\left(H_j|I\right)}.
\end{equation}
The data dependent part is the so called ``Bayes factor'' (BF), and it is defined as:
\begin{equation}
    BF_{ij} \equiv \frac{p \left( D|H_i,I \right)}{p\left(D|H_j, I\right)}.
\end{equation}
In the hypothesis in which, due to our a priori ignorance, all the model's priors are equal, considering the odds ratio is equivalent to considering the Bayes factor.

The case of a model $H_i(\vec{\theta_i})$ dependent on a vector of parameters $\vec{\theta_i}$ is more complicated because we have to compute the marginal likelihood of the model, which corresponds to the likelihood of the model averaged over all possible values of the parameters. For this reason, it is often called average or marginalised likelihood. The marginal likelihood is computed by marginalising over the model's parameters: 
\begin{equation}
\begin{split}
    p\left(D|H_i, I\right) = &\int p( D, \vec{\theta}_i| H_i, I )\, d\vec{\theta}_i \\
    = &\int p(D|H_i, \vec{\theta}_i, I)
     p(\vec{\theta}_i|H_i, I)\, d\vec{\theta}_i, \\
    = &\int \mathcal{L}(\vec{\theta}_i, H_i; D)
    \pi(\vec{\theta}_i|H_i)\, d\vec{\theta}_i,
\end{split}
\end{equation}
where we applied the multiplication law of probabilities, and, in the last equality, expressed the integrand in terms of likelihood and prior probabilities of the parameters. In any real situation, this integral is not solvable analytically and can only be estimated numerically. Nevertheless, in the case where the information content of our measurement is such that the priors can be considered sufficiently vague in the range of the parameters space where the likelihood is sizable, we can consider the prior as a constant and use the Laplace approximation to estimate the integral. This is a very useful approximation to get sense of the different contributions to the marginal likelihood. For simplicity, let's assume the model depends just on a single parameter $\theta_i$, the best fit to the data occurs for $\theta =\hat{\theta}$ (in the following we will use $\hat{x}$ to refer to the best fit value of the quantity $x$), the likelihood is reasonably normal, and the prior is flat over a range $\Delta\theta_i$. Under this assumptions the integral can be approximated as:  
\begin{equation}
\begin{split}
    p\left(D|H_i, I\right) = &\int \mathcal{L}\left(\theta_i, H_i; D\right)
    \pi(\theta_i|H_i)\, d\theta_i, \\
    \simeq &\frac{1}{\Delta\theta_i} \, \sqrt{2\pi} \, \sigma(\hat{\theta_i})\, \mathcal{L}(\hat{\theta_i}, H_i; D).
\end{split} 
\label{eq:bayesapprox}
\end{equation}
The last expression makes it clear that the marginal likelihood corresponds to the likelihood evaluated at the best fit value of $\theta_i$ weighted by a volume factor  corresponding to the parameter's uncertainty. In this oversimplified example, the proportional factor between the global and the best fit likelihood is given by the ratio of two quantities, namely $\Delta\theta_i$ and  $\sigma(\theta_i)$, which control respectively the possible range of $\theta_i$ before and after the data are seen. This factor is often referred to as the Ockham's factor as it naturally penalises a model by a quantity proportional to the necessary complexity of the model to get a good fit with respect to the a priori model complexity; the ideal case where the model is a good one and the a priori variability of the parameter is just enough to fit the data, the Ockham's factor is of order one. The Ockham's factor enters in the BF through the marginal likelihood, and thus when we compare two models we automatically account for the potential difference in their complexity.  

In order to visualise that, let's make a useful example. Let's consider two model:
\begin{itemize}
    \item \emph{Model A}: this models has only one parameter, let's call it $\theta_A$, and the prior over this parameter is a uniform distribution in the range $[\theta_A^1,\, \theta_A^2]$ of width $\Delta\theta_A$, and thus the pdf is:
    \begin{equation}
        p\left(\theta_A |H_A,I\right) = \frac{1}{\Delta\theta_A} 
    \end{equation}
    \item \emph{Model B}: this models has two parameters, $\theta_{B1}$ and $\theta_{B2}$, that are uncorrelated and the prior over this parameters is a uniform distribution of width $\Delta\theta_{B1}$ and $\Delta\theta_{B2}$, respectively:
    \begin{equation}
        p\left(\theta_{B1}, \theta_{B2} |H_B,I\right) =
        p\left(\theta_{B1}|H_B,I\right)
        p\left(\theta_{B2}|H_B,I\right) = 
        \frac{1}{\Delta\theta_{B1}}  \frac{1}{\Delta\theta_{B2}} 
    \end{equation}
\end{itemize}

Assuming normal likelihoods, and using the eq.~(\ref{eq:bayesapprox}), the BF can be estimated as:
\begin{equation}
    BF_{AB} \simeq \frac{p(D|\hat{\theta}_A, H_A, I)}
    {p(D|\hat{\theta}_{B1}, \hat{\theta}_{B2}, H_B, I)}
    \times
    \frac{\sqrt{2 \pi} \sigma (\hat{\theta}_A)}{\Delta\theta_A}
    \bigg/    
    \left(    
    \frac{(2 \pi) \sigma (\hat{\theta}_{B1}) \sigma (\hat{\theta}_{B2})}
    {\Delta\theta_{B1}\Delta\theta_{B2}}
    \right)
        = LR_{AB} \times OF_{AB},
\label{eq:bayesfactor_example}
\end{equation}
where we defined the likelihood ratio $(LR_{AB})$
and the Ockham's factor ($OF_{AB}$) as
\begin{equation}
\begin{split}
    LR_{AB} &= \frac{p(D|\hat{\theta}_A, H_A, I)}{
        p(D|\hat{\theta}_{B1}, \hat{\theta}_{B2}, H_j, I)
        },\\
    OF_{AB} &= \frac{\sqrt{2 \pi} \sigma (\hat{\theta}_A)}{\Delta\theta_A}
    \bigg/    
    \left(    
    \frac{(2 \pi) \sigma (\hat{\theta}_{B1}) \sigma (\hat{\theta}_{B2})}
    {\Delta\theta_{B1}\Delta\theta_{B2}}
    \right)
\end{split}
\end{equation}

The message behind eq.~(\ref{eq:bayesfactor_example}) is that even though a model with more parameters can be more flexible and thus better fit the data producing an higher likelihood, one has to pay the price of having a more complex model.
 In other words: even if the likelihood ratio pushes towards a more complex model which often better fits the data, on the other hand the Ockham's factor penalises it. 

It is important to stress that for parametric models the Bayes factor critically depends on the choice of priors, that, in this case, is represented by the choice of the widths $\Delta\theta_i$. For models with the same number of parameters, especially if the parameters have the same physical meaning, the Ockham's factor should be of order one, and the BF should fairly be insensitive to priors' choice. 

For models with a different number of parameters one has to pay particular attention to avoid introducing biases due to an `unreasonable choice of priors'. For this reason, in the following we will start by comparing models with just one parameter and then move to more complex and realistic scenarios exploring the sensitivity of our conclusions to the prior's choice. 

\section{Details of the {\tt JAGS} MCMC simulation}\label{appendix:JAGS}
The full hierarchical probability model described in Section (\ref{subsec:implementation}) is implemented in {\tt JAGS}. To define the model one uses the {\tt BUGS} language \cite{bugs-paper,bugs-lunn, bugs}, which, for example for the COS model, looks as follows:\\
\begin{lstlisting}[language=R]
 model { 
   for (i in 1:n) {
       y[i] ~ dnorm(mu[i], 1./(sd[i]^2))
       mu[i] <- A * cos(2.*pi*(t[i] - t0)/T)
   }
   # prior on parameters 
   A ~ dunif(0., maxA)
}
\end{lstlisting}
where the {\tt for} loop is over the entries of the data $n$-tuple, {\tt dnorm()} stands for the normal distribution density function whose parameters are the true amplitudes {\tt mu[i]} as given by the functional form of the model, and the standard deviations {\tt sd[i]} given as $\tau_i = 1/\sigma_i^2$. The last line defines the prior pdf for the parameter of interest $A$, which, in this case is given by a uniform distribution in the range $[0,\,{\tt maxA}]$. This model will be fit to time series data described by an $n$-tuple $\{D_i\}=\{t_i, y_i, \sigma_i\}$ of length $n$, containing for each time bin $t_i$ the measurement $y_i$ and the symmetric uncertainty $\sigma_i$ associated to it. 

\subsection{Prior choices}\label{appendix:prior}

\begin{figure}
    \centering
    \includegraphics[width=.48\textwidth]{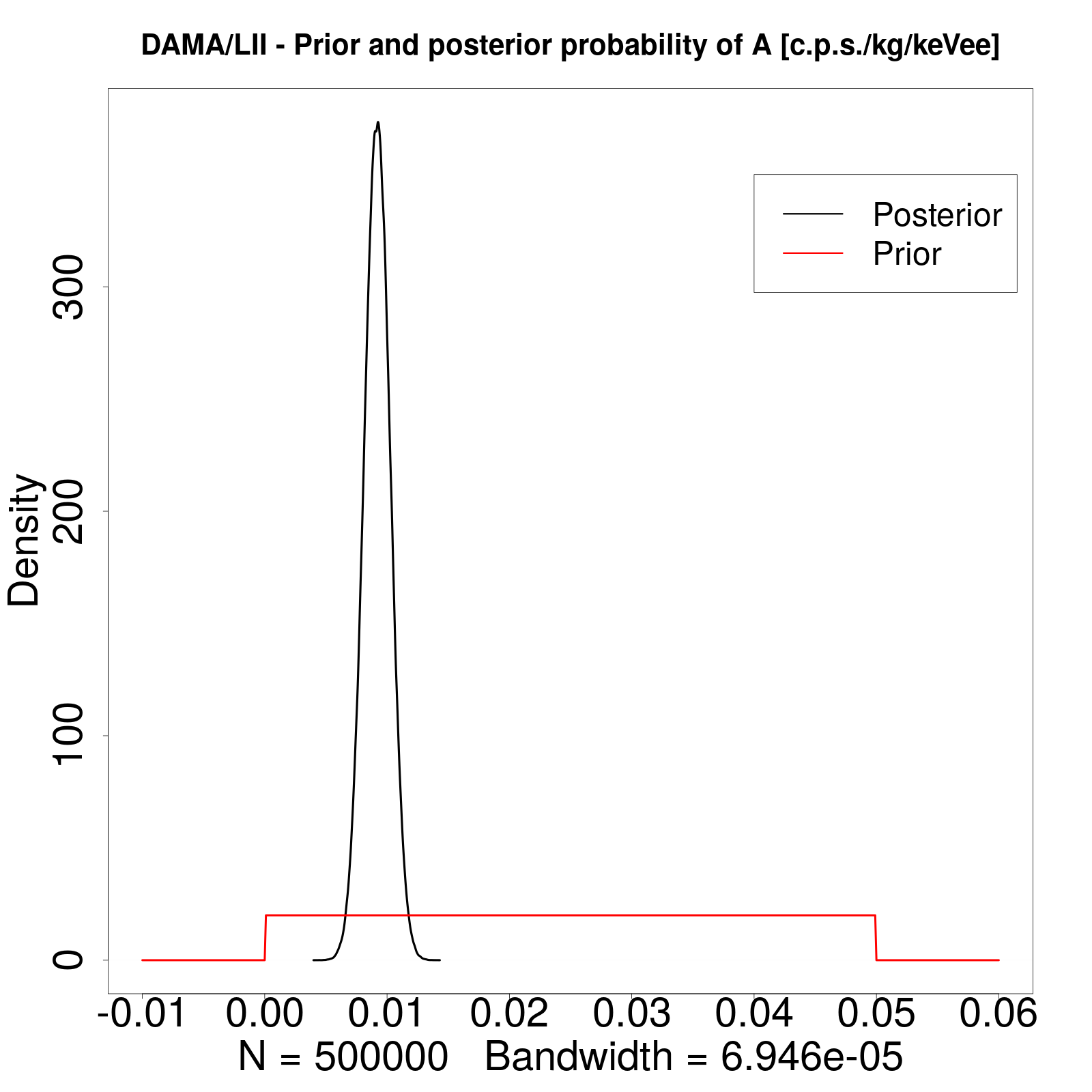}
    \includegraphics[width=.48\textwidth]{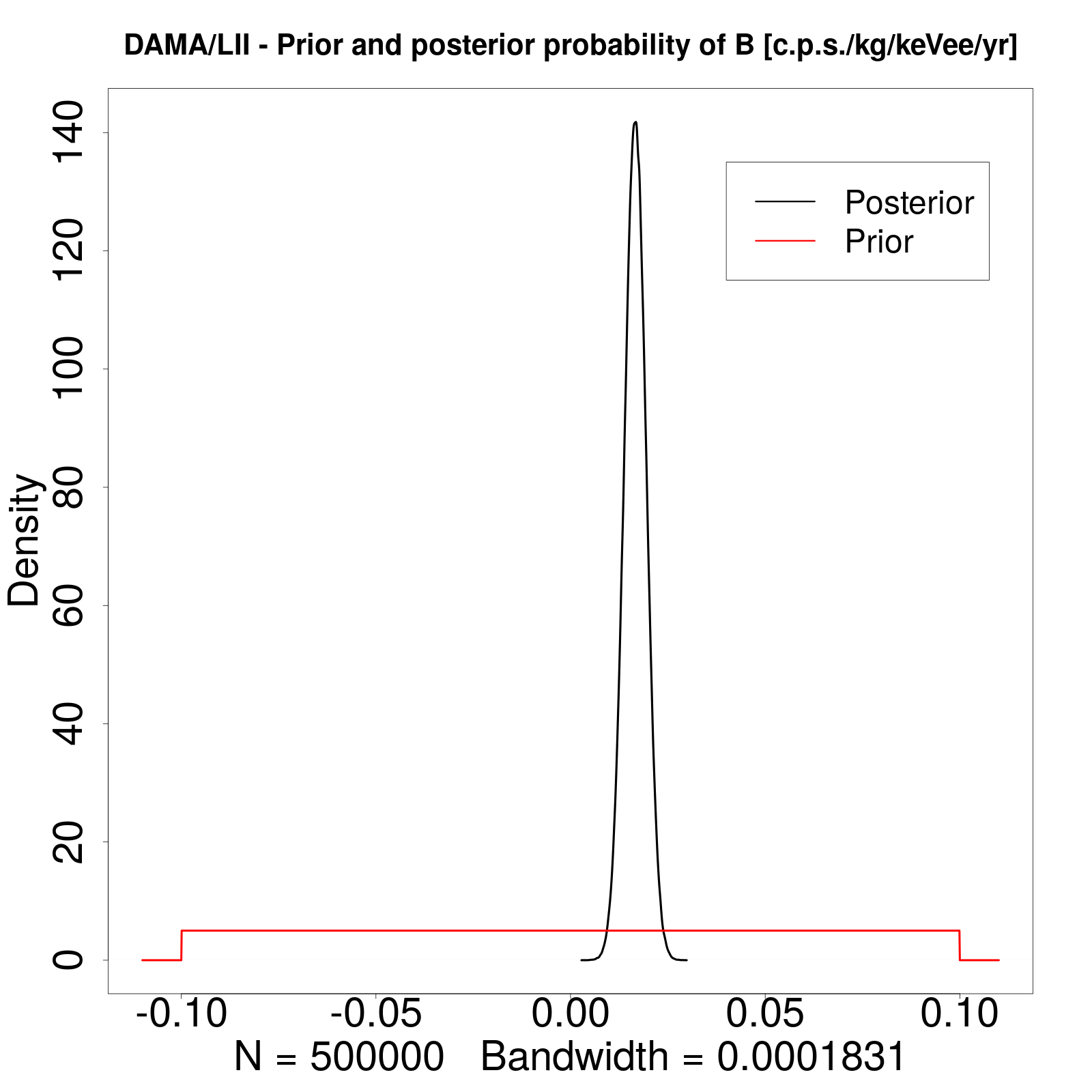}
    \caption{\label{fig:priorpostAB} \em Posterior pdf (black lines) and priors pdf (red lines) for the free parameters $A$ of the COS model (left) and $B$ of the SAW model (right) for the fit of the DAMA/LIBRA Phase II data.}
\end{figure}

\begin{figure}
    \centering
    \includegraphics[width=.48\textwidth]{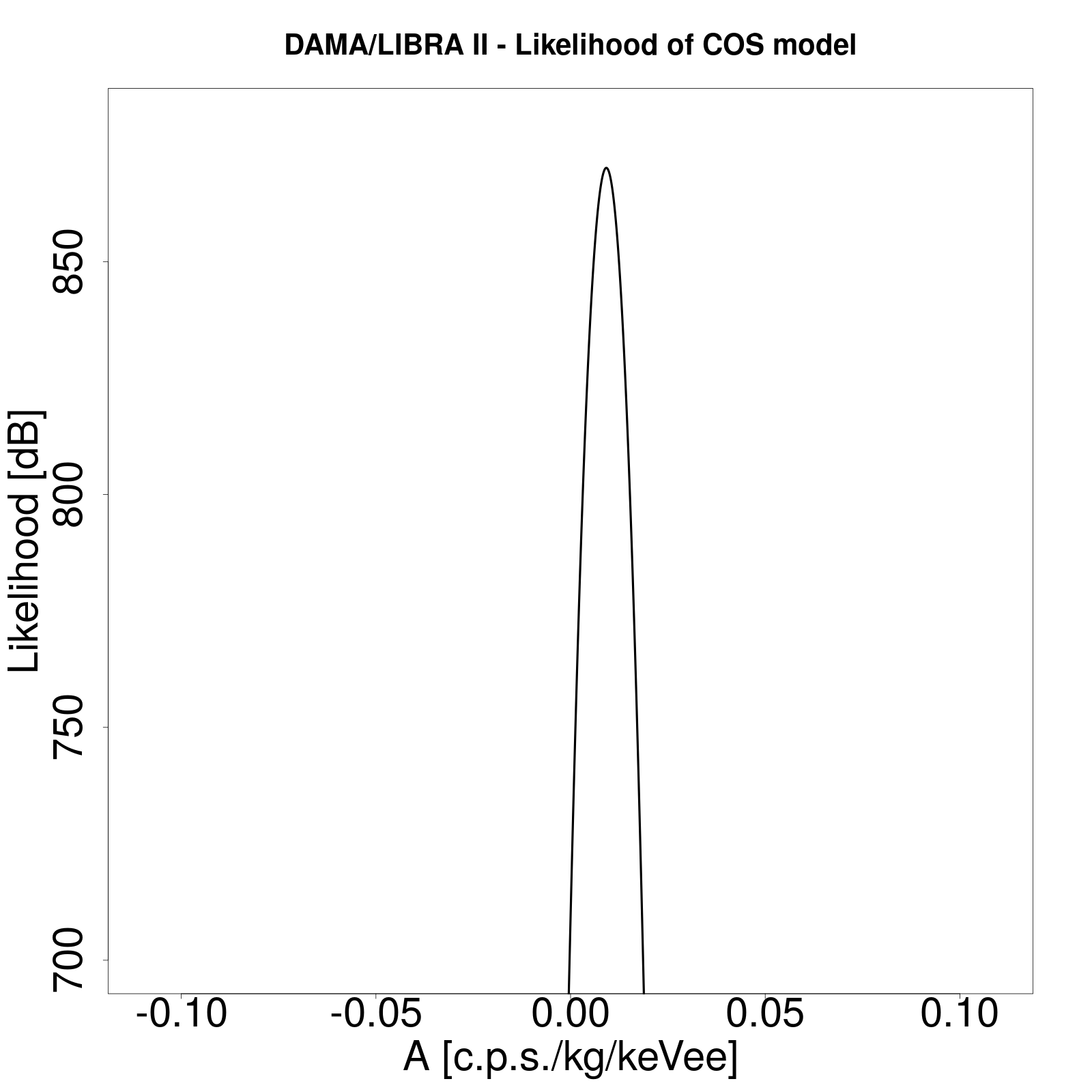}
    \includegraphics[width=.48\textwidth]{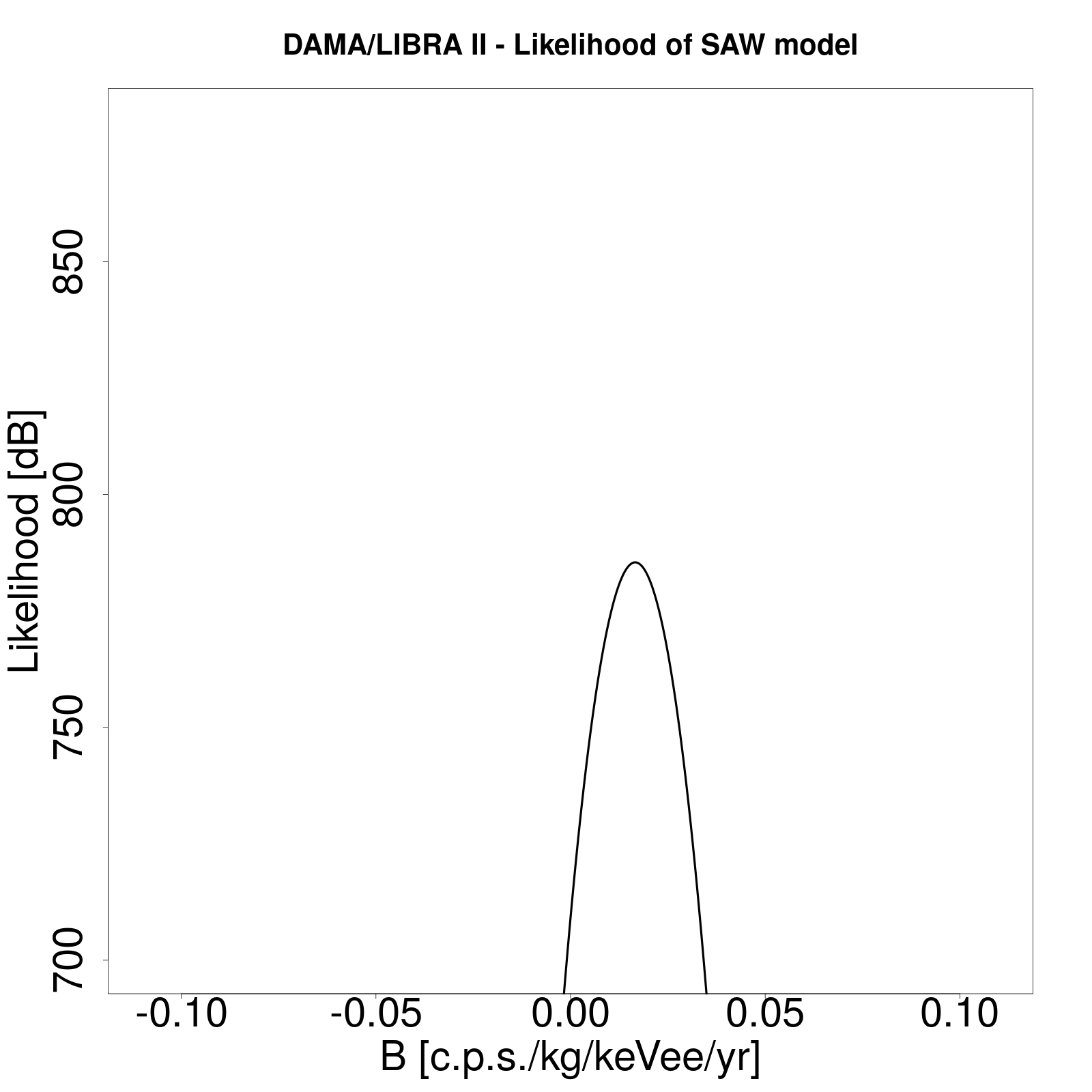}
    \caption{\label{fig:likcossaw} \em Likelihood of the DAMA/LIBRA Phase II data for the COS model (left) and the SAW model (right).}
\end{figure}

\begin{figure}
    \centering
    \includegraphics[width=.48\textwidth]{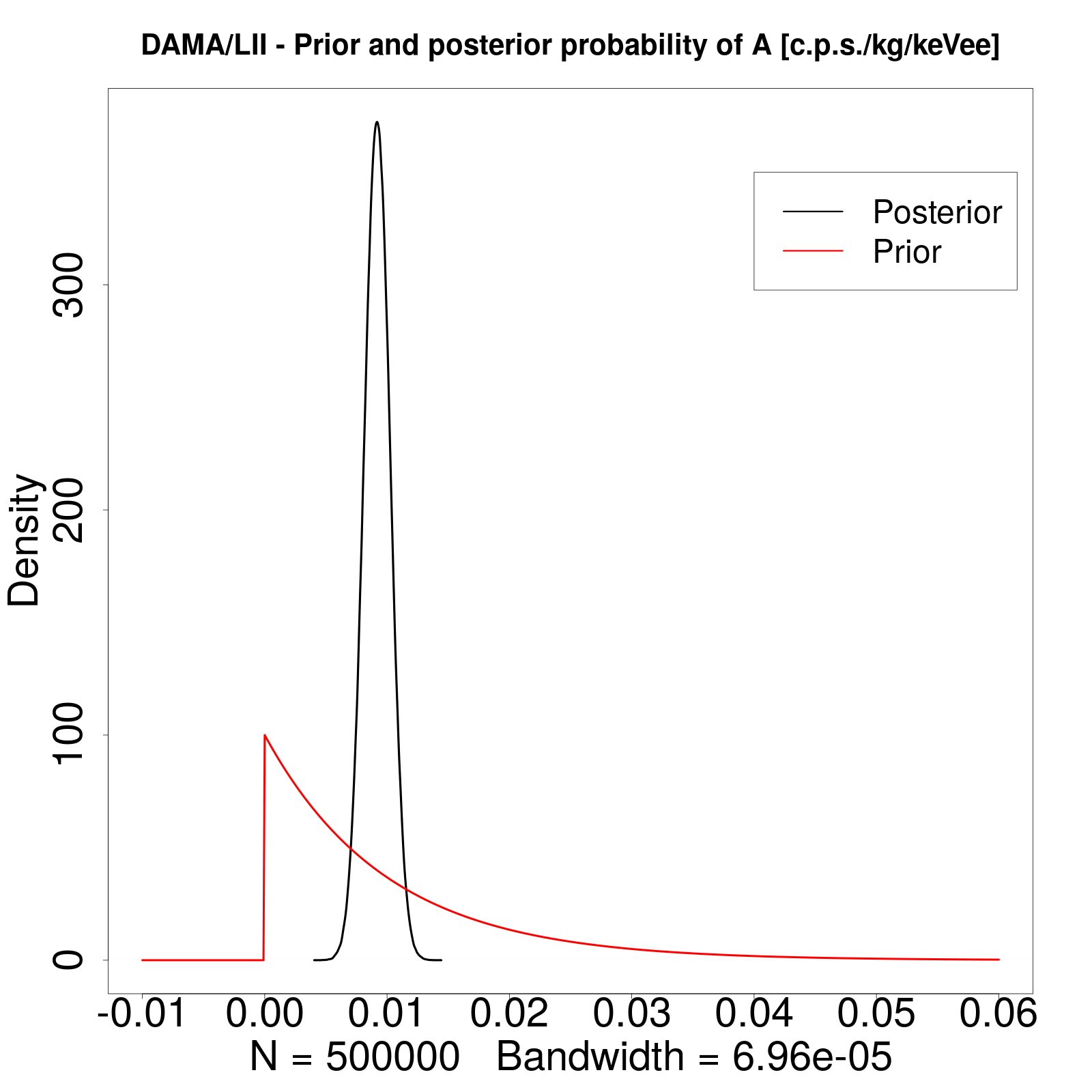}
    \includegraphics[width=.48\textwidth]{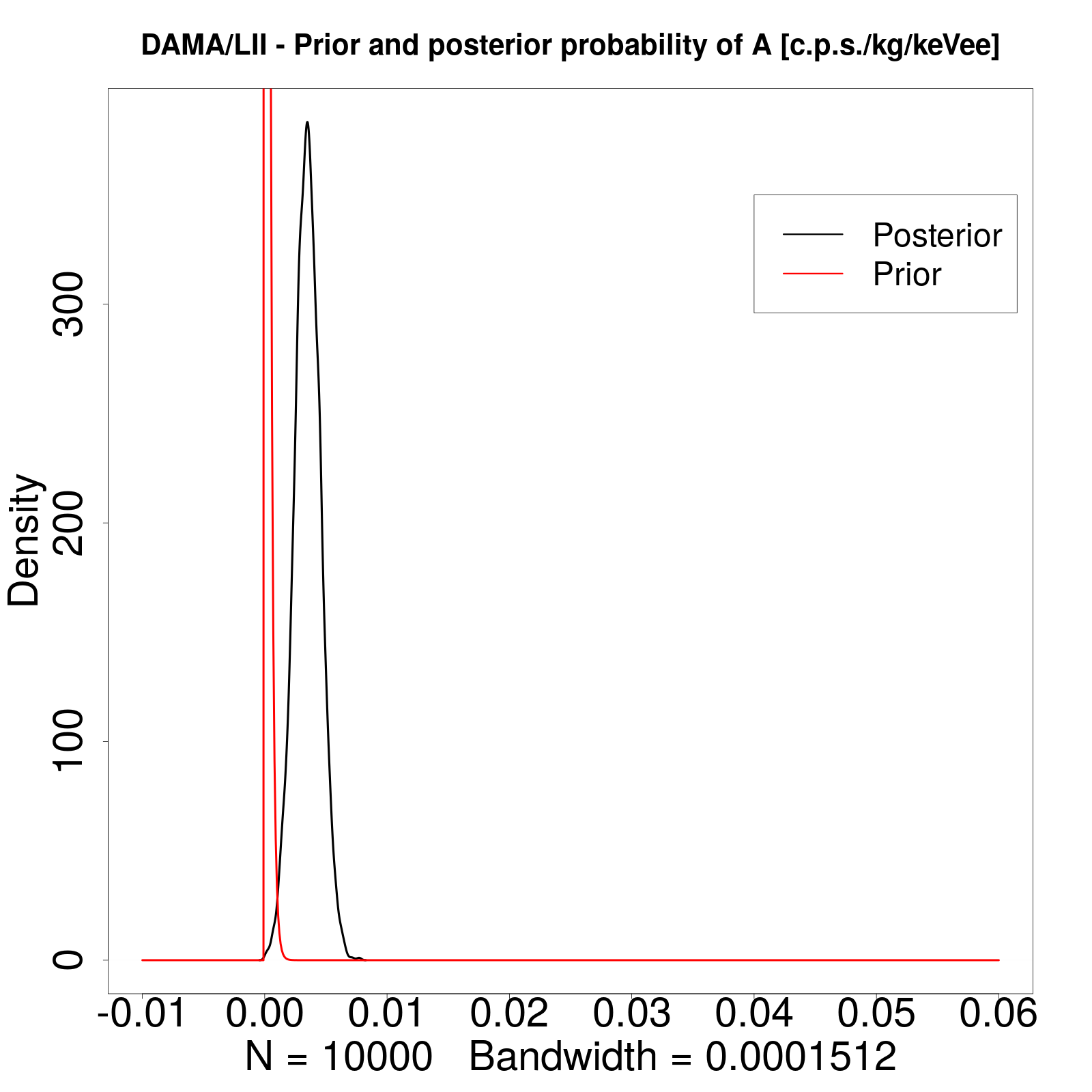}
    \caption{\label{fig:priorpostgamma} \em Posterior pdf (black lines) and priors pdf (red lines) for the free parameter $A$ of the COS model for the fit of the DAMA/LIBRA Phase II data: the priors in these cases are gamma distributions with a shape $\alpha = 1$ and a scale $\theta = 100$ (left) or $\theta=5000$ (right). As expected, the posterior is quite different from the flat prior case of fig.~\ref{fig:priorpostAB} only in the most extreme case (right).}
\end{figure}

The priors used in our analysis are the following:
\begin{itemize}
    \item for all the cases in which there's a cosine modulation the amplitude $A$ has a uniform prior:
    \begin{equation}
        \pi(A) = 
        \begin{cases}
            &\frac{1}{\Delta_A} \quad \text{if}
            \quad 0\leq A \leq \Delta_A \\
            &0 \quad \quad otherwise 
        \end{cases}
    \end{equation}
    with $\Delta_A = 0.05\:cpd/kg/keVee $. One could in general choose a wider prior, but with the upper limit set by the results of the other DM experiments.
    \item for all the cases in which there's a sawtooth contribution, the slope $B$ has a uniform prior:
    \begin{equation}
        \pi(B) = 
        \begin{cases}
            &\frac{1}{2\Delta_B} \quad \text{if}
            \quad -\Delta_B \leq B \leq \Delta_B \\
            &0 \quad \quad otherwise 
        \end{cases} 
    \end{equation}
    with $\Delta_B =0.1\:cpd/kg/keVee/yr$. In this case we allowed $B$ to have negative values in order to take into account a possible slowly decreasing sawtooth. As in the previous case, one could in general choose a wider prior, but with the upper limit on the background set by the knowledge of the DAMA experimental apparatus.
    \item for the case in which we let the period and the phase of the cosine free to vary, the two parameters $T$ and $t_0$ have a uniform prior:
    \begin{equation}
    \begin{split}
        \pi(T) &= 
        \begin{cases}
            &\frac{1}{2 \Delta_T} \quad \text{if}
            \quad \tilde{T}-\Delta_T \leq T \leq \tilde{T} +\Delta_T \\
            &0 \quad \quad otherwise 
        \end{cases} \\
        \pi(t_0) &= 
        \begin{cases}
            &\frac{1}{2\Delta_{t_0}} \quad \text{if}
            \quad \tilde{t_0} -\Delta_{t_0} \leq t_0 \leq \tilde{t_0} + \Delta_{t_0} \\
            &0 \quad \quad otherwise 
        \end{cases} \\
    \end{split}
    \end{equation}
    where:
    \begin{equation}
        \begin{split}
        \tilde{T} &= 1 \: yr \\
        \tilde{t_0} &= 152.5 \: d = 0.418 \: yr \\
        \Delta_T &= \tilde{T}/2 = 0.5 \: yr \\
        \Delta_{t_0} &= \tilde{t_0}/2 = 0.209 \:yr
        \end{split}
    \end{equation}
    In this case, since we would like to study the effect of letting these two parameters free to vary, we chose quite large priors around the expected values for a DM signal.
\end{itemize}

Since in all cases, the likelihoods are quite narrow with respect to the priors we choose, we were able to easily estimate the OF using the Laplace approximation as shown in Appendix~\ref{app:bayesian}.

In all this cases these priors could be different in principle, and we checked, by changing both the width of the priors and their functional form, that the conclusions of our analysis wouldn't change with other reasonable choices. As an example, in fig.~\ref{fig:priorpostAB} we show the posterior
and the prior for the parameters of the COS and SAW models in the DAMA/LIBRA Phase II experiment; for the sake of completeness, in fig.~\ref{fig:likcossaw} we show the likelihoods of the two models in the same dataset. In fig.~\ref{fig:priorpostgamma}, instead, we show, always as an example, how the posterior of the COS parameter in the DAMA/LIBRA Phase II dataset would change with two different choices of priors (a gamma distrubution in both cases): in one case, the prior is significantly different from zero in the region where the likelihood has the maximum, and thus it does not affect significantly the posterior pdf which is comparable with that of fig.~\ref{fig:priorpostAB} (fig.~\ref{fig:priorpostgamma}: left-hand side); in the other case, the prior pdf is so much concentrated in zero that the posterior changes significantly (fig.~\ref{fig:priorpostgamma}: right-hand side).

\subsection{Posterior results}

\begin{figure}[!ht]
    \centering
    \includegraphics[width=.48\textwidth]{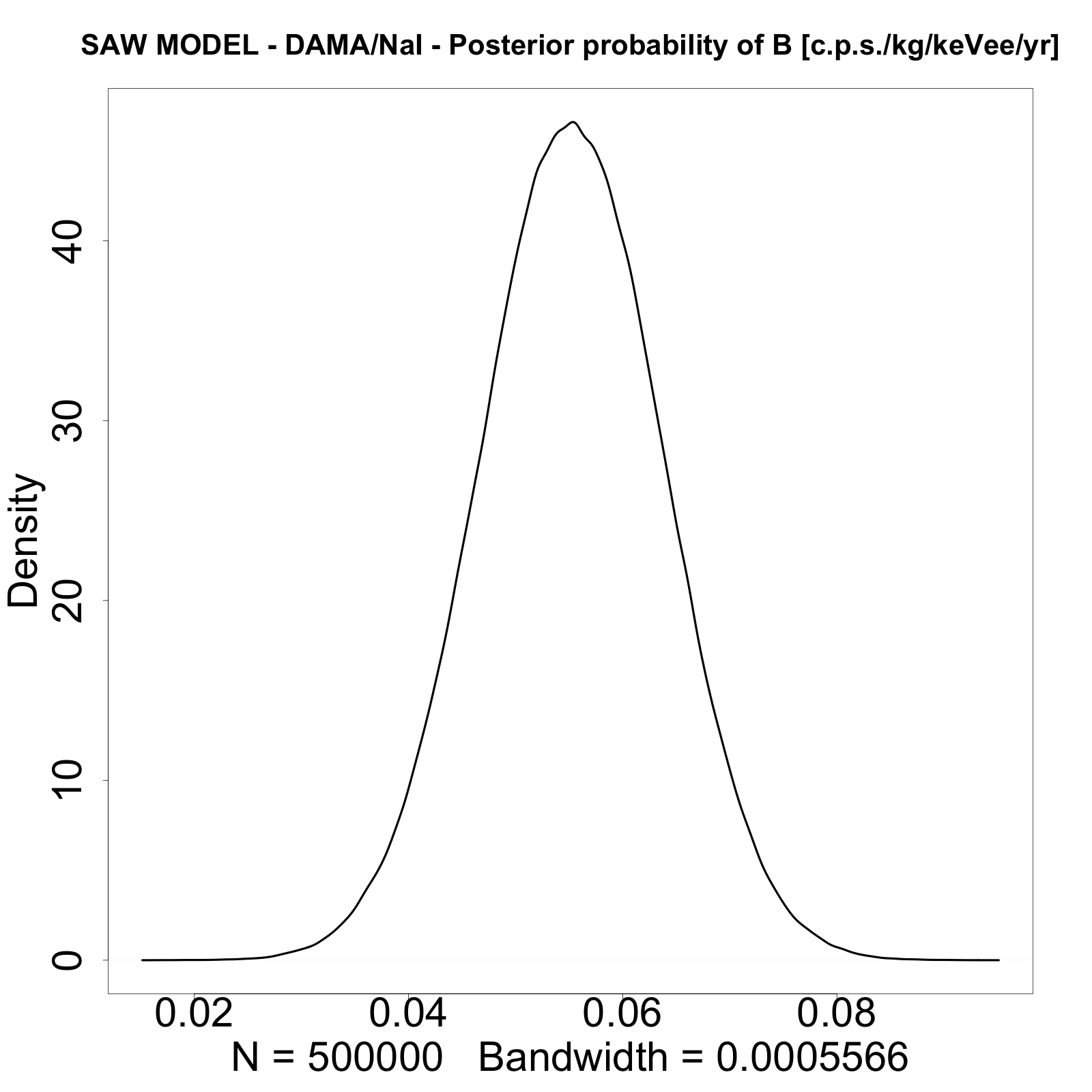}
    \includegraphics[width=.48\textwidth]{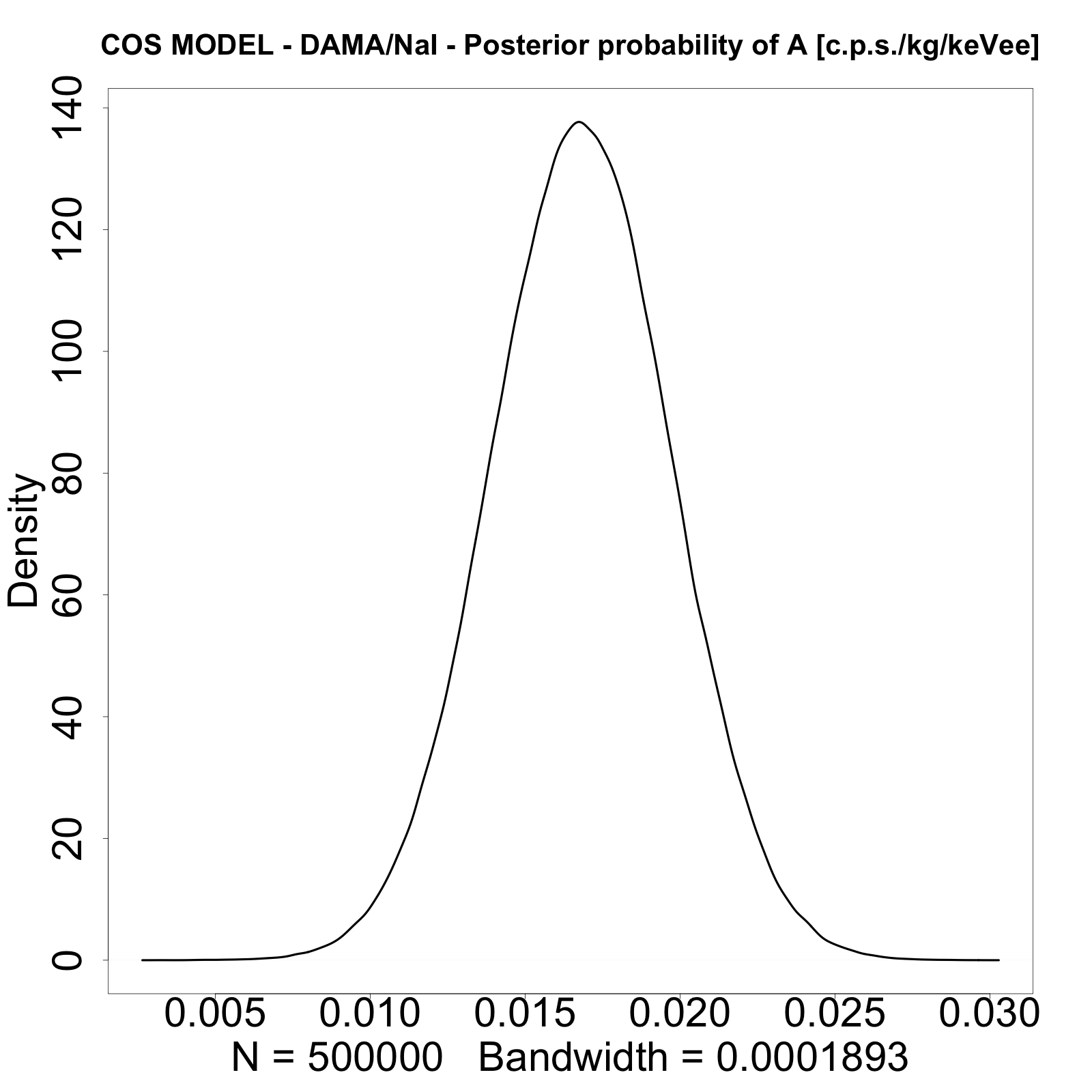}
    \caption{\label{fig:posteriorABNaI} \em Posterior pdfs for the free parameters of the two models for the fit of the DAMA/NaI data. On the left hand side, the parameter $B$ for the SAW model, and, on the right hand side, the parameter $A$ for the COS model. }
\end{figure}

\begin{figure}[!ht]
    \centering
    \includegraphics[width=.48\textwidth]{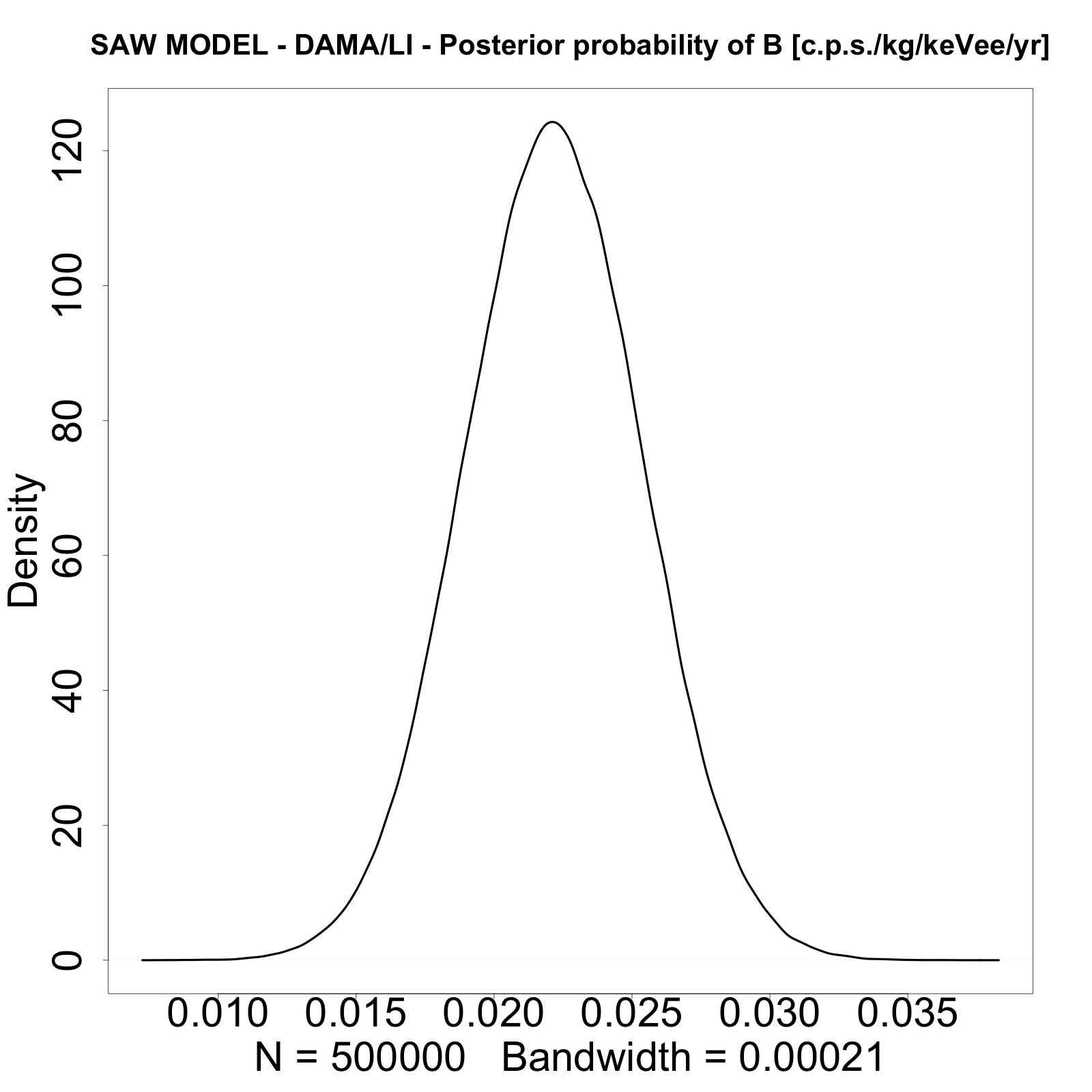}
    \includegraphics[width=.48\textwidth]{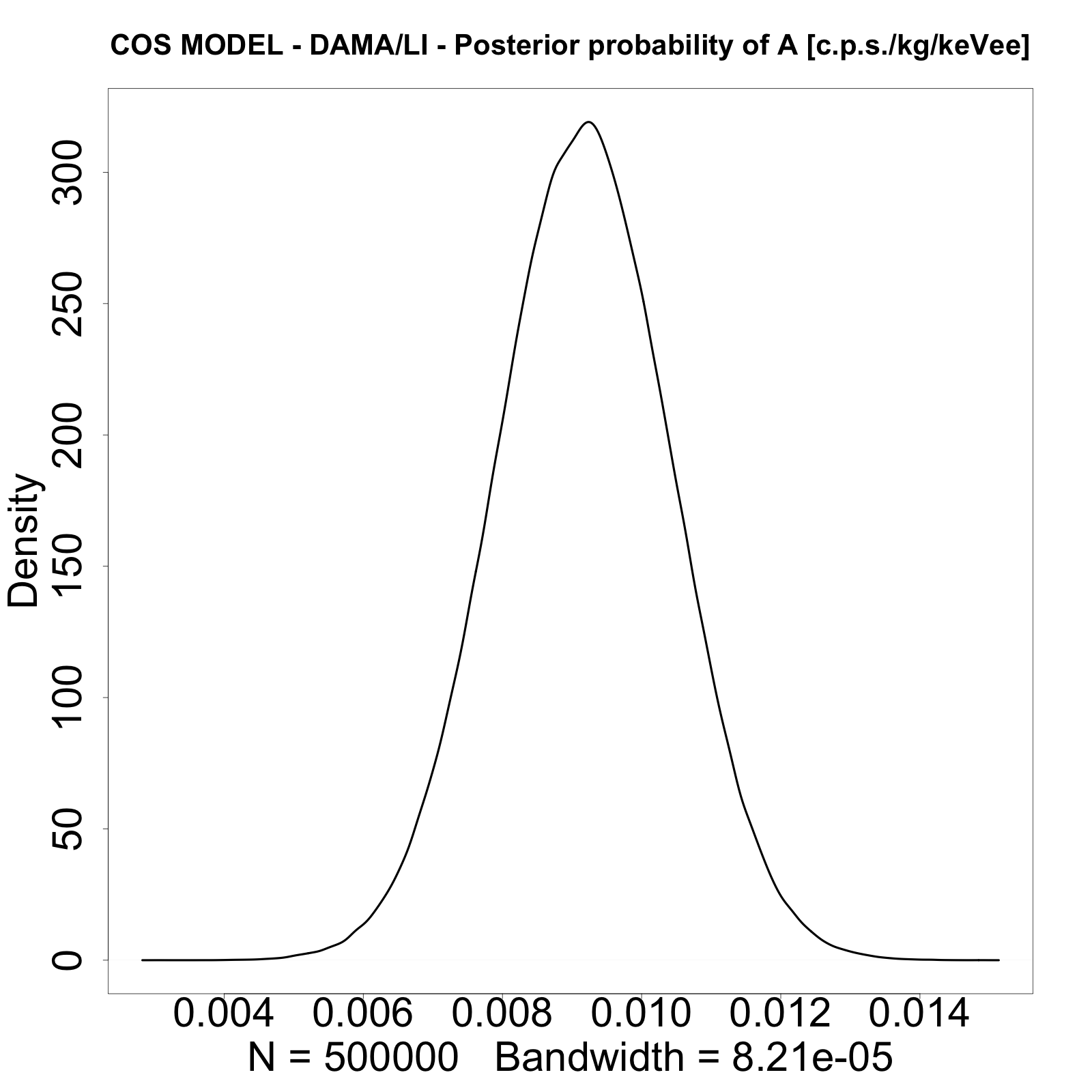}
    \caption{\label{fig:posteriorABLI} \em Posterior pdfs for the free parameters of the two models for the fit of the DAMA/LIBRA I data. On the left hand side, the parameter $B$ for the SAW model, and, on the right hand side, the parameter $A$ for the COS model. }
\end{figure}

\begin{figure}[!ht]
    \centering
    \includegraphics[width=.48\textwidth]{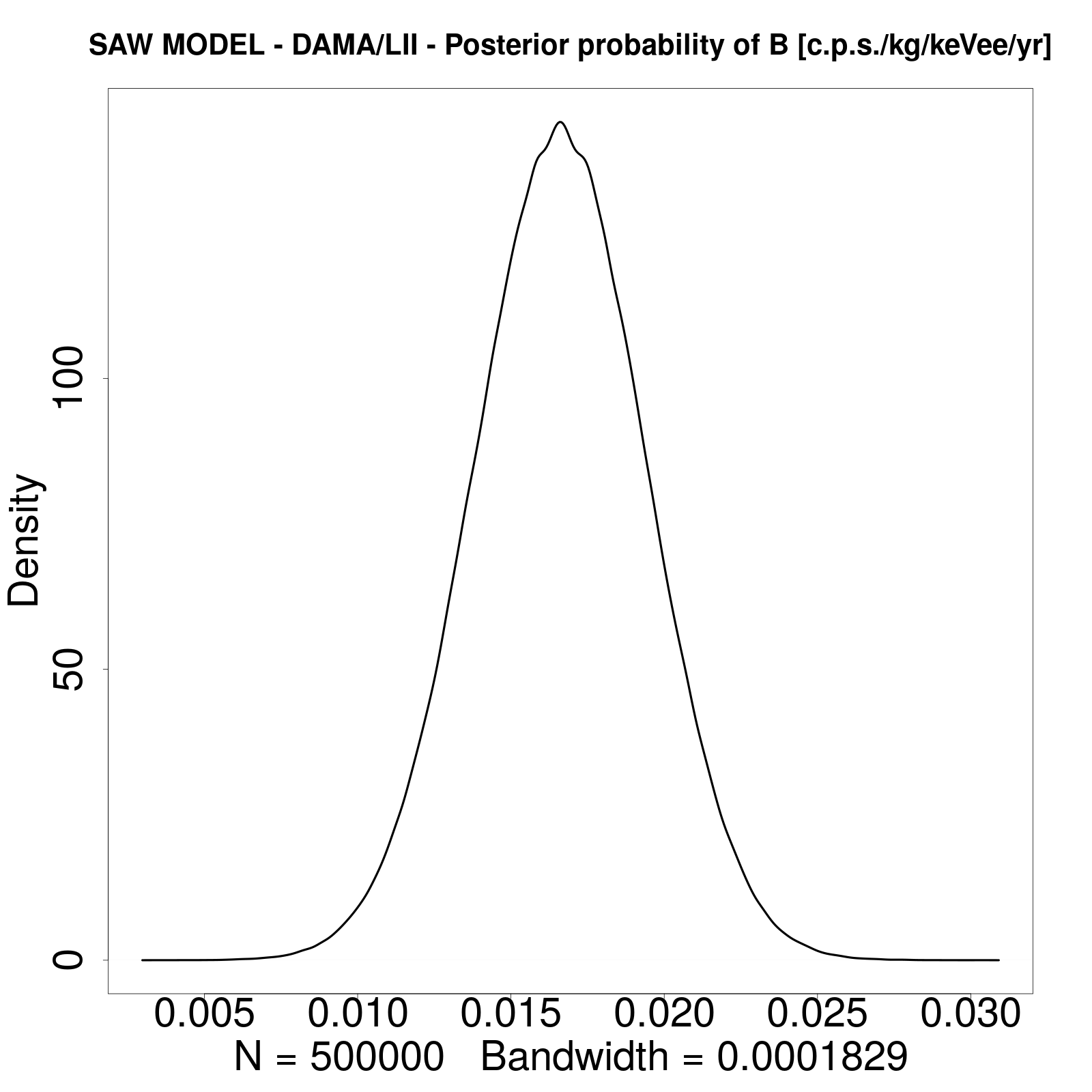}
    \includegraphics[width=.48\textwidth]{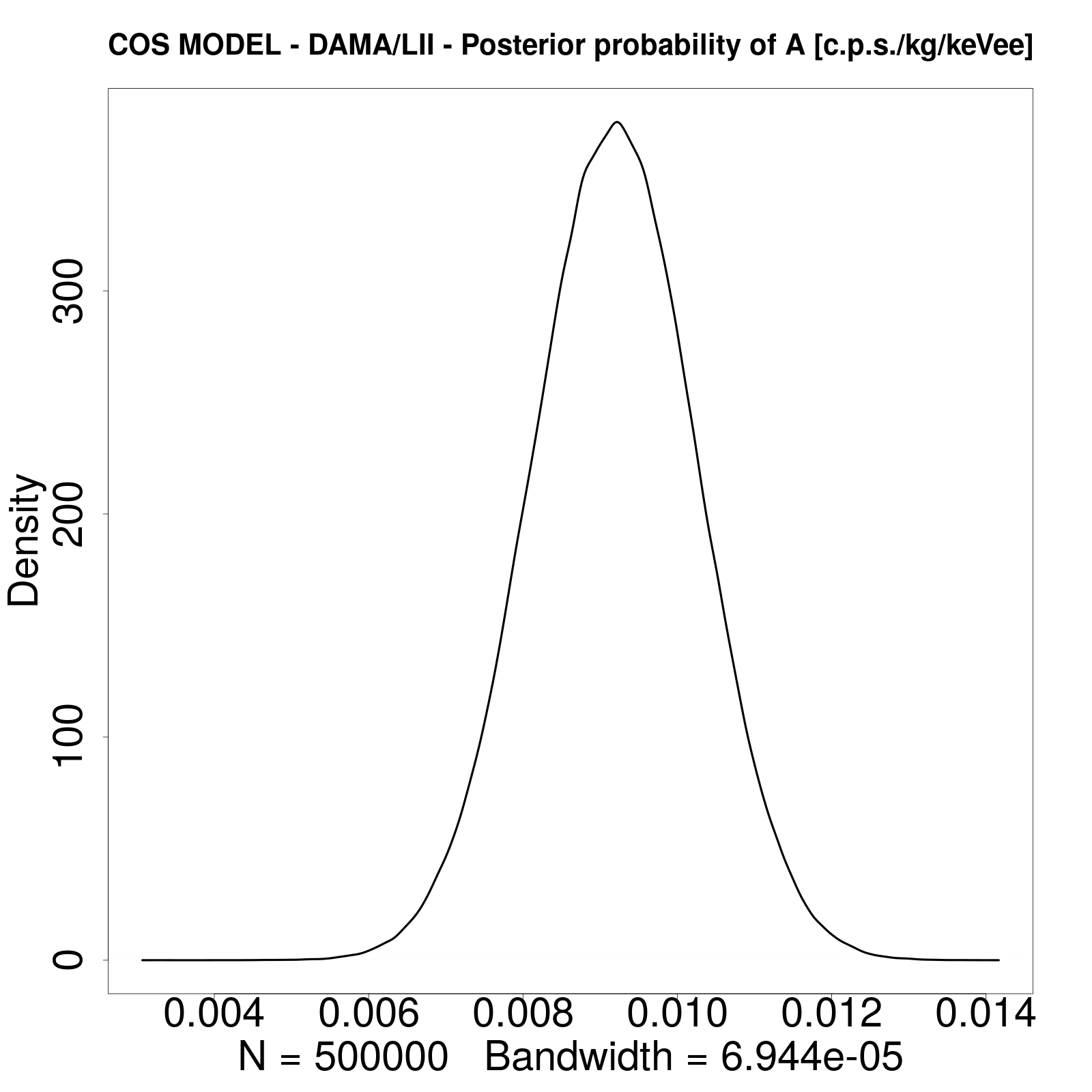}
    \caption{\label{fig:posteriorABLII} \em Posterior pdfs for the free parameters of the two models for the fit of the DAMA/LIBRA II data. On the left hand side, the parameter $B$ for the SAW model, and, on the right hand side, the parameter $A$ for the COS model. }
\end{figure}

\begin{figure}[!ht]
    \centering
    \includegraphics[width=.9\textwidth]{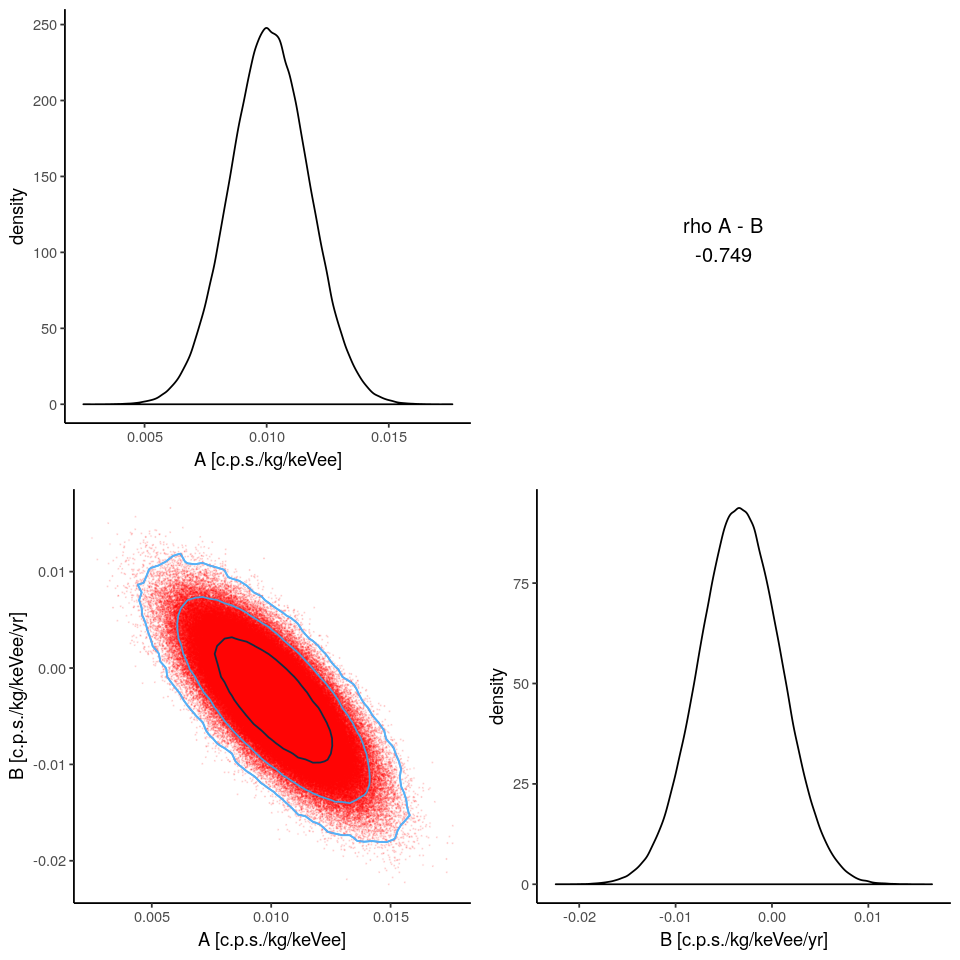}
    \caption{\label{fig:posteriorABLIICS} \em Posterior pdfs for the free parameters of the C+S model for the fit of the DAMA/LIBRA II data. The figure shows  in the lower-left corner the combined pdf reporting also the credible regions at $0.68,0.95,0.997$ probability as black, green, and blue contours respectively. The upper-left and lower-right plots are the marginal pdfs for the parameter A and B respectively. The correlation coefficient is given in the upper-right corner of the figure.}
\end{figure}

\begin{figure}[!ht]
    \centering
    \includegraphics[width=.5\textwidth]{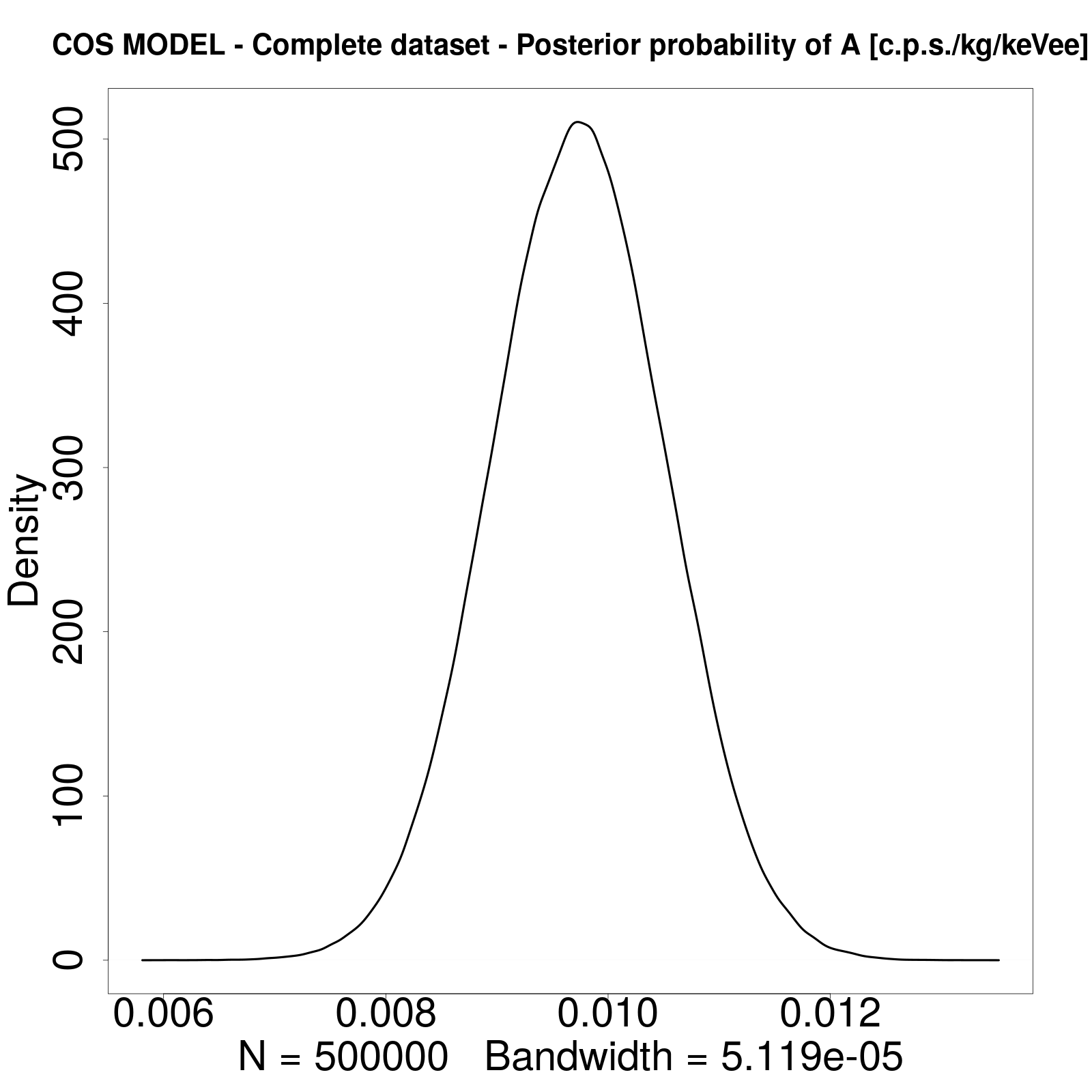}
    \caption{\label{fig:posteriorA123} \em Posterior pdf for the free parameter of the COS model for the fit of the whole (2-6) keVee DAMA dataset.
   }
\end{figure}

\begin{figure}[!ht]
    \centering
    \includegraphics[width=.9\textwidth]{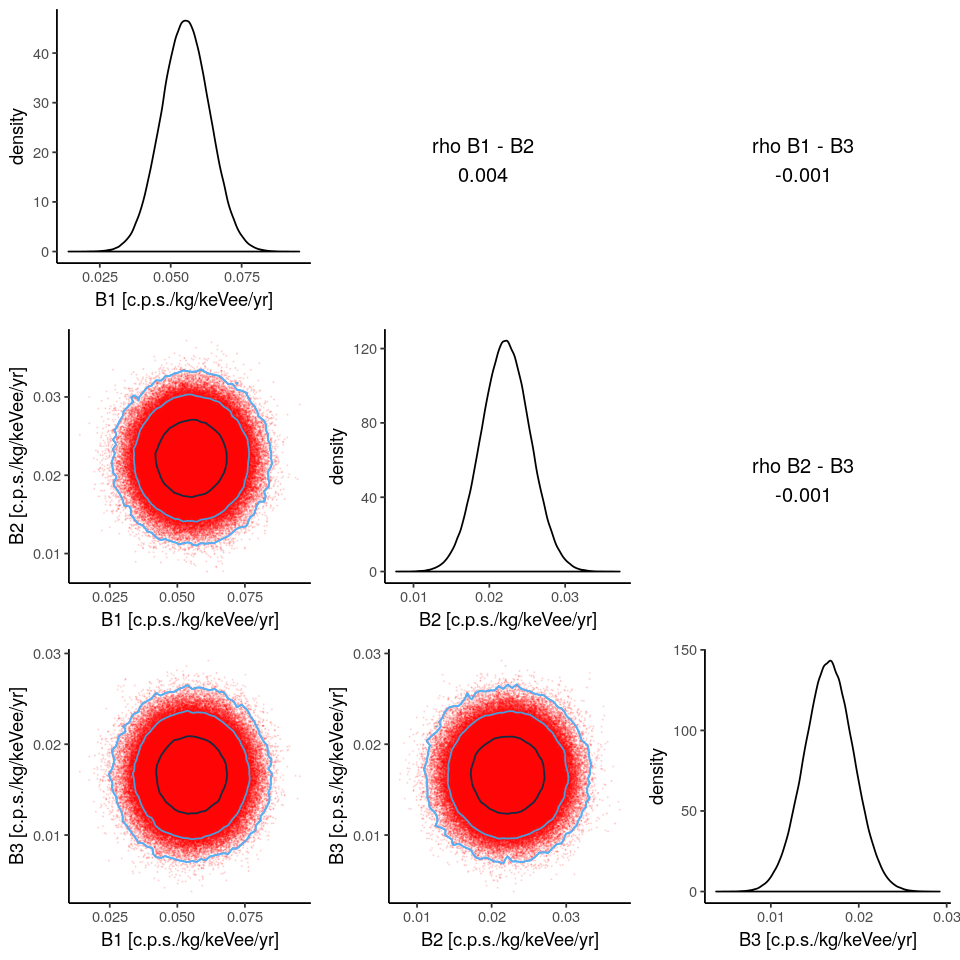}
    \caption{\label{fig:posteriorB123} \em Marginal posterior pdfs for the free parameters of the SAW model with three different slopes for the fit of the whole (2-6) keVee DAMA dataset.
     The 3 plots on the diagonal of the figure are the uni-dimensional pdfs of each single free parameter obtained by marginalising on the all the others. The 3 bi-dimensional pdfs in the bottom-left corner of the figure give the marginal pdfs of each pair of parameters obtained by marginalising on the other. The plots show also the credible regions at $0.68,0.95,0.997$ probability as black, green, and blue contours respectively. The correlation coefficients are given in the upper-right corner of the figure.}
\end{figure}

\begin{figure}[!ht]
    \centering
    \includegraphics[width=.9\textwidth]{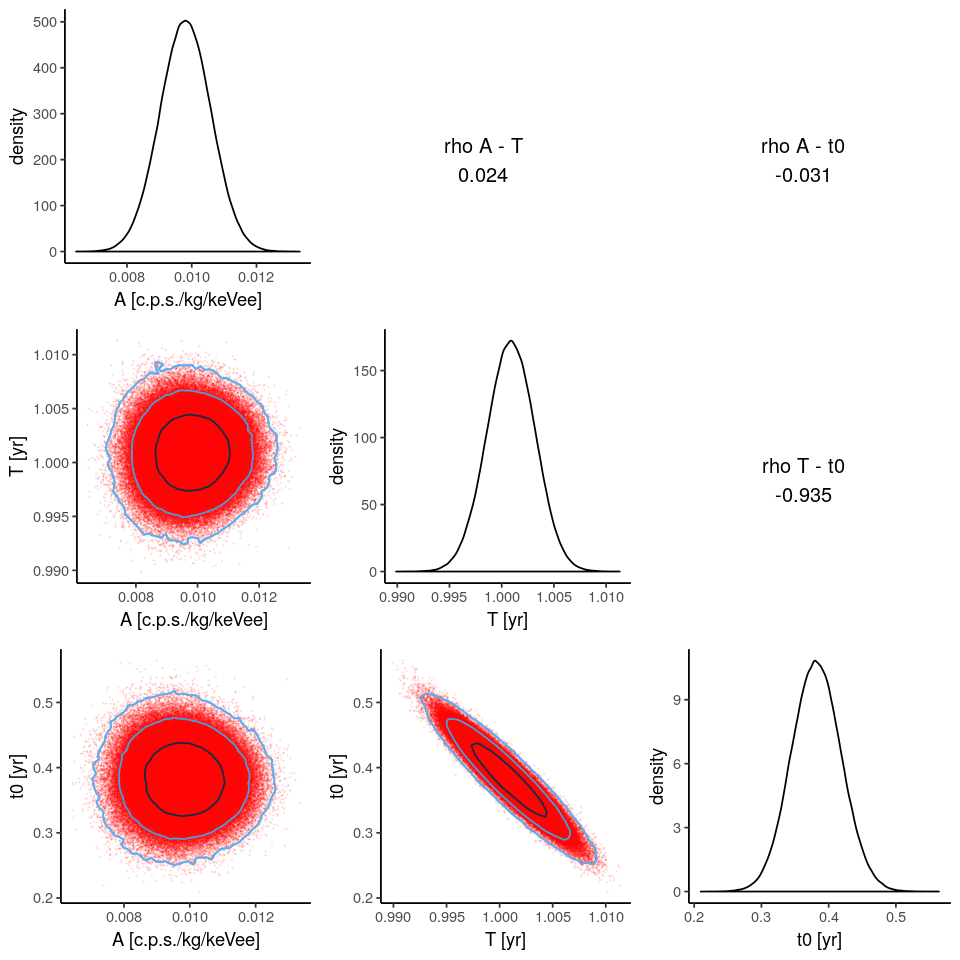}
    \caption{\label{fig:posteriorcosine} \em Marginal posterior pdfs for the free parameters of the cosine model for the fit of the whole (2-6) keVee DAMA dataset.
     The 3 plots on the diagonal of the figure are the uni-dimensional pdfs of each single free parameter obtained by marginalising on the all the others. The 3 bi-dimensional pdfs in the bottom-left corner of the figure give the marginal pdfs of each pair of parameters obtained by marginalising on the other. The plots show also the credible regions at $0.68,0.95,0.997$ probability as black, green, and blue contours respectively. The correlation coefficients are given in the upper-right corner of the figure.}
\end{figure}

The results of our fits, given in terms of posterior pdfs, are showed in fig.~\ref{fig:posteriorABNaI}, \ref{fig:posteriorABLI}, \ref{fig:posteriorABLII}, \ref{fig:posteriorABLIICS}, \ref{fig:posteriorA123}, \ref{fig:posteriorB123}, \ref{fig:posteriorcosine}. In all cases, the MCMC chains reach equilibrium after few steps showing a stable trace.  The numerical error on the results sampled with $\mathcal{O}(10^5)$ steps is negligible. As one can see from all the figures listed here, all the posteriors are reasonably normal, as could be expected from the assumption of normal likelihoods and uniform priors. In particular:
\begin{itemize}
    \item fig.~\ref{fig:posteriorABLIICS} shows the expected anti-correlation between $A$ and $B$ of the C+S model and, in addition, indicates that $B$ is consistent with zero.
    \item fig.~\ref{fig:posteriorB123} shows, as expected,  that the three $Bi$ slopes of the sawtooth for the SAW model  are not correlated when fitted on the three stages of the experiment.
    \item fig.~\ref{fig:posteriorcosine} shows the posterior distributions of the parameters $A$, $T$ and $t_0$ of the cosine model when we let the period and the phase free to vary. As expected $T$ and $t_0$ are strongly anti-correlated, while we don't see any remarkable correlation between the other parameters. 
\end{itemize}

Finally, in fig.~\ref{fig:posteriorcomplete} we show the result of a complete fit on all the (2-6) keVee dataset, obtained using a C+S model with three different slopes in the three different stages and treating the period and the phase of the cosine modulation as free parameters, and, in fig.~\ref{fig:posteriorcomplete4}, we display the result of the same fit, but fixing the period and the phase of the cosine modulation to their expected values ($T = 1\:yr$ and $t_0 = 152.5 \: d$). 


\clearpage


\end{document}